\documentclass[journal]{IEEEtran}
\ifCLASSINFOpdf
\else
\fi
\usepackage{graphics}
\usepackage{epsfig}
\usepackage{times}
\usepackage{amsmath}
\usepackage{wrapfig}
\usepackage{subfigure}
\usepackage{array}
\usepackage{multirow}
\usepackage{multicol}
\usepackage{rotating}
\usepackage{balance}
\usepackage{cite}
\usepackage{color}
\usepackage{cancel}
\usepackage[table]{xcolor}
\usepackage[noend]{algpseudocode}
\usepackage{algorithm}
\usepackage{hyperref}

\def \etal {et al.}
\def \ie {i.e.,}

\begin{document}
%
\title{Curriculum Learning for Speech Emotion Recognition from Crowdsourced Labels}
%
%
%

\author{Reza~Lotfian,~\IEEEmembership{Student Member,~IEEE,}
        and~Carlos~Busso,~\IEEEmembership{Senior~Member,~IEEE}
\thanks{\IEEEcompsocthanksitem R. Lotfian and C. Busso are with the Erik Jonsson School of Electrical \& Computer Engineering, The University of Texas at Dallas, Richardson TX 75080.\protect\\
E-mail: reza.lotfian@utdallas.edu, busso@utdallas.edu}
}

%
%



\maketitle

\begin{abstract}
This study introduces a method to design a curriculum for machine-learning to maximize the efficiency during the training process of \emph{deep neural networks} (DNNs) for speech emotion recognition. Previous studies in other machine-learning problems have shown the benefits of training a classifier following a curriculum where samples are gradually presented in increasing level of difficulty. For speech emotion recognition, the challenge is to establish a natural order of difficulty in the training set to create the curriculum. We address this problem by assuming that ambiguous samples for humans are also ambiguous for computers. Speech samples are often annotated by multiple evaluators to account for differences in emotion perception across individuals. While some sentences with clear emotional content are consistently annotated, sentences with more ambiguous emotional content present important disagreement between individual evaluations. We propose to use the disagreement between evaluators as a measure of difficulty for the classification task. We propose metrics that quantify the inter-evaluation agreement to define the curriculum for regression problems and binary and multi-class classification problems. The experimental results consistently show that relying on a curriculum based on agreement between human judgments leads to statistically significant improvements over baselines trained without a curriculum. 
\end{abstract}


%

\section{Introduction}
\label{sec:intro}

\IEEEPARstart{E}{motion} recognition can play an important role in advanced \emph{human computer interaction} (HCI) applications that are more aware of the users. Recently, there has been a number of studies showing the feasibility of emotion recognition in different applications, opening novel research directions. For example, in \emph{human-robot interaction} (HRI), a robot capable of recognizing and expressing facial expressions can more effectively communicate and interact with humans \cite{Bartlett_2003}. Emotion recognition systems have also been used in designing interactive games \cite{Szwoch_2015,Obaid_2008} and tutoring systems \cite{Litman_2004,Litman_2006} to teach social interactions to children with autism spectrum disorders \cite{Abirached_2012}.

While previous studies have made important advances, emotion recognition from speech still faces many challenges. The most important challenges are generalizing the models across datasets \cite{Busso_2013, Schuller_2010_2, Zhang_2011_2}, accounting for individual differences \cite{Vogt_2006,Schuller_2005,Busso_2013_2}, and recognition of subtle expressions \cite{Scherer_2003,Mower_2009}.

Recently deep learning techniques have enabled advances in many areas of speech processing. A barrier of using deep learning in speech emotion recognition is the lack of large training datasets. While tasks such as \emph{automatic speech recognition} (ASR) have access to datasets with several hours of speech samples collected from multiple speakers, most affective datasets contain only tens or hundreds of subjects displaying a small range of emotions. The limited resources make it extremely difficult to train models that generalize well within and across datasets. Therefore, it is important to utilize the limited available training resources in the most efficient way. 

Inspired by language learning in children, Elman \cite{Elman_1993} demonstrated that a neural network is able to better learn a task when the training data is sequentially presented from simple to complex. When the order is random, the network may fail to converge to an optimal set of parameters. This training approach is referred to as curriculum learning and has been successfully used in word representations in natural language processing \cite{Turian_2010}, working memory based problem solving tasks \cite{Krueger_2009}, controlling robots \cite{Sanger_1994, Baranes_2013}, and executing computer programs \cite{Zaremba_2014_2}. More recently, Bengio \etal \cite{Bengio_2009_2} demonstrated that curriculum learning results in better generalization and faster learning on synthetic vision and word representation learning tasks. Interestingly, this approach has not been used on speech emotion recognition. We hypothesize that curriculum learning can be effectively employed to deal with the limited training resources available in this area. We also hypothesize that using a curriculum leads to models that learn more general emotion cues before learning more subtle speaker dependent cues, improving the generalization of the models. For these reasons, this work explores curriculum learning for speech emotion recognition. Emotion recognition has attributes which make it a good candidate for applying curriculum learning training schemes. The main evidences are the complex nature of the problem and the way we learn to perceive emotions, which happens gradually from infantry to adulthood.

This study consistently demonstrates that prioritizing less difficult samples at the beginning of the training process of a \emph{deep neural network} (DNN) leads to significant improvements for speech emotion recognition systems. We suggest different methods to detect challenging examples which are only used in the more advanced phases of the training process. By adjusting the learning rate of the model in decreasing order, we reduce the influence of difficult samples in setting the parameters of the classifier. The key observation to build the proposed curriculum is that ambiguous samples for humans are also ambiguous for computers. We have observed in our previous work that classifiers trained and tested with samples consistently annotated by human evaluators, achieved higher accuracies than classifiers training with more ambiguous samples \cite{Busso_2017}. This observation suggests that the level of difficulty for each sentence can be estimated by considering the inter-evaluators' agreement. We propose different implementations for this idea using individual annotations available for each sentence. 
We exhaustively evaluate this approach by considering three different formulations for speech emotion recognition. For attribute-based emotional descriptors (e.g., arousal, valence and dominance), we consider regression models to predict the label values. We also consider binary classification after dichotomizing the classes into low versus high values (e.g., low valence versus high valence). For categorical emotions (e.g., happiness, sadness, and anger), we consider a multi-class classification problem. We observe improvement in the performance in all three problems when the training process follows the right curriculum. This approach is better than training with all the samples in one pass. We observe that the best curriculum is based on the item difficulty derived with the minimax conditional entropy criterion suggested by Zhou \etal \cite{Zhou_2014,Zhou_2015}.

The main contributions of this work are two folds. First, to the best of our knowledge, this is the first study on speech emotion recognition using curriculum learning. Second, we propose a novel method to estimate the difficulty of the training examples using individual evaluations. We derived different metrics to quantify the disagreement between evaluators, proposing alternative curriculums for each of the machine-learning formulations considered in this study. The use of curriculum learning leads to consistent performance improvements, demonstrating the benefits of the proposed approach. 

The rest of this paper is structured as follows. Section \ref{sec:related} reviews previous studies on speech emotion recognition that are related to this study. The section also discusses the implementations of curriculum learning in other machine learning applications. Section \ref{sec:method} describes our proposed method to apply curriculum learning to speech emotion recognition. Section \ref{sec:experiment} includes experiments that show the benefit of using the proposed curriculums for this task. Section \ref{sec:conclusion} concludes this study, giving new directions for future research in this area.

\section{Background}
\label{sec:related}

This section reviews studies on speech emotion recognition and the process of annotating emotional labels, which provide the information needed to design the proposed curriculum learning framework (Sec. \ref{ssec:Emorec}). The section also discusses the general idea of curriculum learning and how it has been applied to different problems in previous studies (Sec. \ref{ssec:curlrn}). The section also reviews methods to estimate the difficulty of the tasks without knowing the ground truth for the tasks using minmax conditional entropy framework (Sec. \ref{ssec:minmax}). We apply this method to estimate the difficulty of recognizing emotions conveyed in each speech sample.

\subsection{Speech Emotion Recognition}
\label{ssec:Emorec}

The problem of detecting human emotions in speech, like any supervised machine-learning problem, relies on training examples. Early attempts for collecting emotional databases relied on actors to collect expressive behaviors with predefined emotional category \cite{Petrushin_1999,Dellaert_1996, Busso_2004}. This approach led to satisfactory results when testing under the same database, but failed to perform well under naturalistic situations \cite{Batliner_2000}. In everyday spontaneous conversations, emotions are subtly expressed, creating expressive displays that are not well represented by portrayed behaviors provided by actors, especially if the recordings correspond to read sentences \cite{Batliner_2000,Devillers_2005}. Therefore, the classifiers trained with acted corpora do not generalize well in real life situations. To address this issue, researchers are collecting speech samples under more naturalistic conditions, either by indirectly eliciting target emotions \cite{Busso_2008_5,Busso_2017} or recording spontaneous speech \cite{Devillers_2006,Grimm_2008}. In these scenarios, the emotional content of the speech samples are not determined, and emotional labels need to be assigned using perceptual evaluations \cite{Cowie_2000,Burmania_2016_2}. 

Depending on the subtlety of the emotions, raters often disagree on the label selected for a speech segment, indicating that this is a difficult cognitive task. Recent studies have even leveraged the disagreement between evaluators to train more robust classifiers using soft labels \cite{Lotfian_2017,Fayek_2016}. In our previous work, we demonstrated that classifiers trained with more ambiguous samples achieved lower classification performance that classifiers trained with samples that are consistently evaluated by raters \cite{Busso_2017}. This result suggested that ambiguous samples for evaluators are also ambiguous samples for speech emotion classifiers. Lee \etal \cite{Lee_2011,Lee_2009} proposed to learn categorical emotions by using a hierarchical framework, where easy problems are solved before resolving more difficult problems. They obtained improved performance over multi-class classifiers without this hierarchical approach. These studies suggest that curriculum learning can be an appropriate framework for speech emotion recognition by more effectively using the limited size of emotional databases.

\subsection{Curriculum Learning}
\label{ssec:curlrn}

Most supervised machine learning methods take a one-pass learning approach, where all the training data is used to build the models.
Studies have argued that this is not the best approach, suggesting that a classifier should learn basic patterns from clear examples, leaving harder examples for later training stages \cite{Elman_1993}. The idea of learning simple patterns before complex patterns is usually called curriculum learning. This idea is implemented during the training process by first presenting examples that can be easily recognized, after determining the difficulty level of the samples.  

Bengio \etal \cite{Bengio_2009_2} studied different approaches to implement this idea, showing that a curriculum that introduces the training data from easy to hard can lead to better local minima when training a classifier with a non-convex criterion, which is commonly the case when training DNNs. This training approach results in better generalization and speed-up the convergence. Defining the training objective by giving higher weight to the easier samples is equivalent to solving a smoothed version of the target criterion. The easiest optimization problem $C_0(\theta)$ shares the same parameters ($\theta$) as the target problem $C_1(\theta)$, capturing its coarse structure. The parameter $\lambda$ for the intermediate problems of the curriculum, $C_{\lambda}$,  has to monotonically increase based on the difficulty of the samples from $0$ in the easiest problem ($C_0$) to $1$ in the target problem ($C_1$). A training example $z$ receives a weight to manipulate the difficulty of a problem. Designing a curriculum requires to adjust the weight parameter $W_{\lambda}(z)$ such that the difficulty of $C_{\lambda}$ monotonically increases with $\lambda$. Bengio \etal \cite{Bengio_2009_2} considered discrete numbers of $\lambda$ and also discrete weights ($0$ or $1$) for the training examples ($z$) at each step. As the value of $\lambda$ increases by changing $W_{\lambda}(z)=0$ to $W_{\lambda+\epsilon}(z)=1$, they gradually introduced more difficult samples to the training pool, therefore, increasing the overall difficulty of the problem. They showed the effectiveness of this approach on different problems by training DNNs, comparing the results with a framework using the common one-pass learning approach.

If a natural order to quantify the difficulty of the task is available, the design of the curriculum becomes straightforward. In shape recognition of objects, Bengio \etal \cite{Bengio_2009_2} suggested to start with basic shapes. They increased the difficulty of the task by varying the object position, size, and orientation. They also changed the gray levels of the foreground and background in the image. In language modeling, they predicted the most likely word given the previous words in a sentence in grammatically correct English. For this problem, they controlled the difficulty of the training samples by controlling for the vocabulary size \cite{Bengio_2009_2}. When quantifying the difficulty of the training set is not clear, studies have proposed alternative methods. An intuitive approach for supervised task is to train a classifier using the conventional one-pass learning approach, evaluating the models on the training set. A sample is considered difficult if it falls on the incorrect side of the classifier's hyperplane \cite{Graves_2017}. The distance of the sample to the hyperplane can also be informative, where samples located close to the hyperplane in the feature space are considered more difficult examples than samples that are far from the hyperplane. Gui \etal \cite{Gui_2017} use this method as a baseline to build a curriculum for facial expression analysis. They compared this method to a curriculum built by relying on the intensity of the expressed emotions. In language model, this model-driven approach to build the curriculum can be implemented by assigning the difficulty level based on whether the model is able to predict the next word \cite{Bengio_2009_2}. These methods are vulnerable to over-fitting, since the difficulty of the samples is obtained by testing the training examples with a classifier trained on the same data.

Instead of relying on the performance of previously trained classifiers to extract the difficulty information, this study proposes to rely on the labels provided by human annotators by measuring their inter-evaluator agreement. The motivation behind this framework is that, in recognizing emotions, humans clearly outperform existing artificial intelligence solutions, and, therefore, it is reasonable to assume that a curriculum based on a metric of reliability across evaluators may provide a suitable criterion for this task. The challenge is that we usually do not have access to the confidence level of human annotators. Therefore, we rely on the implementation of the minmax conditional entropy method proposed by Zhou \etal \cite{Zhou_2015, Zhou_2014} to indirectly estimate the level of difficulty of a sample, creating an appealing curriculum for speech emotion recognition.

\subsection{Minmax Method for Crowdsourced Labels}
\label{ssec:minmax}

The proposed criteria to define curriculum is based on the minmax method. Recognizing emotions from speech is a subjective task where different evaluators often disagree on the perceived emotion \cite{Steidl_2005}. Therefore, researchers tend to rely on ratings provided by evaluations with different reliability, aggregating the results to achieve consensus labels for the train set. If the reliability of all the raters are identical, the straightforward approach to find consensus judgment from multiple dissident annotations is to find the average for numerical labels or use the majority vote rule for categorical labels. In case the individual's reliability is known, we can rely on weighted majority consensus by giving higher weights to the annotations from reliable raters. In practice, the reliability of an individual worker is unknown. This problem is particularly important for evaluations conducted using crowdsourcing. When ratings are collected through crowdsourcing evaluations, some workers are accurate, while others are less reliable. Some raters are only interested in the payments, limiting their effort as much as possible, providing very noisy annotations. 

To correct the bias of unreliable raters in the overall decision, Dawid and Skene \cite{Dawid_1979} introduced an \emph{expectation maximization} (EM) algorithm to simultaneously estimate the bias of annotators and the label classes in an iterative process. A key part of the Dawid and Skene method is to estimate a latent probabilistic confusion matrix for generating labels for each worker. The off-diagonal elements of the matrix represent the probabilities that the worker mislabels an item from one class as another. The diagonal elements correspond to his/her accuracy for each class. The assumption is that the performance of a worker characterized by his/her confusion matrix stays the same no matter which tasks they are assigned. This assumption is not accurate in many labeling tasks, where some items are more difficult to label than others. For example, a worker is more likely to mislabel a difficult item than an easy one. Moreover, an item may be easily mislabeled with another class due to the ambiguity in the sample. Therefore, a fair evaluation of the raters should separately compensate for the difficulty of each task. Zhou \etal \cite{Zhou_2015} developed a minimax conditional entropy method to jointly infer the task difficulty, raters' bias, and the labels of the samples. Our work uses the difficulty measure found in this \emph{minmax entropy} (ME) approach to design the curriculum for speech emotion recognition. This method is explained in more details in Section \ref{sssec:Expert_disagree}.

\section{Methodology}
\label{sec:method}

This section discusses the motivation for using curriculum learning in speech emotion recognition problems (Sec. \ref{ssec:motivation}). It introduces three formulation for speech emotion recognition used to evaluate the proposed curriculum policies (Sec. \ref{ssec:defproblem}). The section also describes the alternative policies for generating the difficulty measure from crowdsourcing labels to build the curriculum (Sec. \ref{ssec:curriculum}).

\subsection{Motivation}
\label{ssec:motivation}

A candidate problem for curriculum learning	 should rely on non-convex optimization \cite{Bengio_2009_2}. Emotion recognition is a good candidate because of its complex nature \cite{Marsella_2010}. A person takes years to master the essential skills to recognize emotion \cite{Zeidner_2003}. Infants start with limited capabilities to recognize emotions, developing with time more sophisticated representations of the structure of emotions \cite{Volling_2002}. Due to the gradual increase of expertise in emotion recognition, psychologists measure the abilities to perceive affects as an important indicator of human emotional intelligence at different ages \cite{Mayer_1999}. This step-by-step nature of the process of acquiring the capability to perceive emotions suggests that curriculum learning can be an effective method for training speech emotion classifiers.

The first step to establish a curriculum for learning emotions by machines is to quantify the difficulty of the training examples. Since there is no explicit way to determine the difficulty of  the emotional content conveyed on sentences, the difficulty has to be indirectly estimated. In many tasks, the true labels exist (e.g., transcriptions in speech, presence or absence of an object in an image), so the difficulty can be set according to the proportion of raters providing a wrong answer. Such a gold-standard is not available in spontaneous emotional sentences, as the perception of emotion varies across listeners. During the annotation of categorical emotions, evaluators are commonly asked to choose the most relevant emotional classes for a given speech sample. Some sentences have ambiguous emotions where more that than one class may be appropriate (e.g., frustration and anger) \cite{Lotfian_2017}. For attribute-based annotations, evaluators may also disagree when assigning absolute scores to the emotional content of the sentences. This work exploits different policies to build a curriculum for the application of speech emotion recognition. We compare this approach with the one-pass method where all the training samples are equally treated (uniform policy) without curriculum. The results confirm our hypothesis, giving statistically significant differences in the classification results.

\subsection{Formulation of Machine Learning Problems}
\label{ssec:defproblem}
Before we introduce the proposed curriculum policies, it is important to define the machine-learning  problems considered in this study, as their implementation are slightly different across problems. We define three machine-learning problems to evaluate the role of curriculum learning in speech emotion recognition: regression prediction of emotional attributes, binary classification of emotional attributes, and multi-class classification of emotional categories.

\subsubsection{Regression of Dimensional Emotions}
\label{sssec:regproblem}

The first task is to predict the emotional dimension scores for the attributes arousal (calm versus active), valence (negative versus positive) and dominance (weak versus strong), using the training set (one regressor per emotional attribute). Each sentence is annotated by multiple annotators, where the ground truth is the average across the annotations. We use the \emph{concordance correlation coefficient} (CCC) metric to measure the performance of the regressors on the testing set.

\subsubsection{Binary Classification of Dimensional Emotions}
\label{sssec:binproblem}
The second problem is a binary classification task where we split the samples into two classes based on their attribute values (e.g., low valence versus high valence). We use their median value such that the classes are balanced. For the test set, we use the same threshold derived from the training set so the classes are not necessarily balanced. Although this formulation introduces artificial classes (see discussion on Mariooryad and Busso about dichotomizing continuous labels \cite{Mariooryad_2017}), this approach is commonly used in studies on emotion recognition \cite{Schuller_2011_3,Wollmer_2009,Rahman_2012}. The performance of the classification problem is measured using the F-score metric, which combines the mean precision and recall rates for low and high classes.

\subsubsection{Classification of Categorical Emotions}
\label{sssec:catproblem}
The third problem corresponds to a multi-class problem where we recognize the categorical class assigned to each sample. The labels for training and testing the classifiers are assigned using the majority vote rule. Samples without agreement are not used for training of testing the system (i.e., the samples are removed from the corresponding sets). The performance metric for this task is also the F-score estimated from the average precision and recall rates across emotional categories.

\subsection{Curriculum Policies for Emotional Speech}
\label{ssec:curriculum}

The task difficulty for each sample depends on the subjective evaluations conducted by multiple annotators and the machine-learning tasks. This section focuses on explaining the proposed curriculum policies for the formulations for speech emotion recognition described in Section \ref{ssec:defproblem}. We suggest three general criteria to design the curriculum for training the classifiers: Criterion 1 is a basic curriculum that relies on the results of a pre-trained classification or regression model. Criterion 2 creates the curriculum by considering the disagreement between evaluators. Criterion 3 also considers the disagreement between evaluators, taking into account the reliability of the annotators.

\subsubsection{Criterion 1: Error of Predicted Label}
\label{sssec:errlabel}

The first criterion uses pre-trained models to determine the difficulty order in the curriculum. First, we train a classifier using all the samples available for training. The models are tested on the same training set. Then, we compare the predicted class with the actual label, defining specific rules to quantify the difficulty of sample $i$ for each machine learning problem, denoted by $d_i$. 

For regression problems, the difficulty of a sample is defined using Equation \ref{eq:confreg} by estimating the distance between the predicted value ($y'_i$) and the ground-truth ($y_i$). The easiest training samples have $d_i=0$, where the pre-trained regression model successfully predicts the actual value without error.  

\begin{equation}
d_i = |y_i - y'_i|
\label{eq:confreg}
\end{equation}

For binary (emotional attributes) and multi-class (categorical emotions) tasks, we define the difficulty of a sentence based on the classification results and the confidence of the classifier. For training sample $i$ with true label $y_i$ and feature vector $\mathbf{x_i}$, the pre-trained classifier predict the label $y'_i$. For the samples that are correctly classified (i.e., $y_i = y'_i$), the difficulty is defined as the confidence of the classifier on the predicted classes (with negative sign). For samples that are incorrectly classified (i.e., $y_i \neq y'_i$), we consider the confidence in the incorrect class as the difficulty metric (Eq. \ref{eq:confclass}). 

\begin{equation}
d_i = \begin{cases}
 -P(y_i= {y}'_i |\mathbf{x_i}), \text{if } y_i = y'_i \\
 P(y_i= {y}'_i |\mathbf{x_i}), \text{if } y_i \neq y'_i
\end{cases}
\label{eq:confclass}
\end{equation}

With this measure, an easy sample has a smaller $d_i$ than a more difficult sample. The level of confidence $P$ is inferred from the classifier's output. For a DNN with softmax function in the output layer, the maximal neuronal response of the softmax layer can be used as a measure of confidence \cite{Geifman_2017}.

The metric $d_i$ for regression (Eq. \ref{eq:confreg}) and classification (Eq. \ref{eq:confclass}) considers the consensus label, which is generated by aggregating the answers from all the evaluators. The variations between individual evaluations have no effect on the difficulty measure, and, therefore, the curriculum.

\subsubsection{Criterion 2: Disagreement Between Annotators}
\label{sssec:disagreement}

The second criterion relies on finding the level of disagreement between annotators for each sentence. Intuitively, annotators will have higher agreement on samples with clear emotional content (\ie \space easy samples), and lower agreement for more emotionally ambiguous samples (\ie \space difficult examples). For regression problem, a metric of disagreement across evaluators is the variance of the scores provided to a sample. Therefore, $d_i$ is defined as: 

\begin{equation}
d_i = \frac{\sum\limits_{j=1}^{N} (y_{ij}-\bar{y_i})^2	}{N_i}
\label{eq:disagreereg}
\end{equation}

\noindent where $y_{ij}$ is the label assigned by annotator $j$ for sample $i$, $\bar{y_i}$ is the average across all the annotations available for sample $i$, and $N_i$ is the number of annotations available for sample $i$. 

For categorical emotions, $d_i $ quantifies how popular is the emotional class with more votes by finding the ratio between annotators selecting that class and the total number of annotators:

\begin{equation}
d_i = \frac{\sum\limits_{j=1}^{N} [y_{ij}=\hat{y_i}]}{N_i}
\label{eq:disagreecateg}
\end{equation}

\noindent where $\hat{y_i}$ is the consensus label for sample $i$ using the majority vote rule. In the binary problems for attribute-based annotations, the dichotomized labels are obtained by splitting the average scores using media split. We also use Equation \ref{eq:disagreecateg} to estimate $d_i$.

\subsubsection{Criterion 3: Minmax Conditional Entropy Inference}
\label{sssec:Expert_disagree}
The third criterion for curriculum learning considers the disagreement across evaluators by modeling the level of expertise of the raters. The previous criterion assumes that all the annotators are equally reliable (Sec. \ref{sssec:disagreement}). In reality, annotators have different level of expertise. For example, some raters are unfamiliar with the concept of emotional attributes, easy distracted, or inconsistent with their judgments. Therefore, we should consider the expertise of the raters to better determine whether the disagreement in the emotional labels is due to poor raters or the difficulty of the samples. A fair assessment process should jointly estimate the task difficulty and annotator's skills. The \emph{item response theory} (IRT) \cite{Lord_1980} is a method to model the probability of a correct answer to a given item by a person with a specific ability level. It uses latent characterization of individuals and items as predictors of observed responses. This model relies on item discrimination, item difficulty, and the probability that an individual with very low ability correctly answers a question. Since the true labels of the items are not available in our case, the most likely correct response need to be estimated in addition to the item difficulty and the workers ability.

To address this problem, Zhou \etal \cite{Zhou_2014,Zhou_2015} employed minmax conditional entropy method subject to constraints to encode the observation for ordinal \cite{Zhou_2014} and categorical \cite{Zhou_2015} labels. In their formulation, the objective is to aggregate crowdsourced labels for a set of items annotated by a group of workers. The input is the observed label ${\tilde{y}_{ij}}$ which is the label selected for item $i$ by annotator $j$. The goal is to estimate the unobserved true labels $y_i$ from the noisy workers' labels (i.e., estimating the probability of the item belonging to each class $Q(Y_i=c)$ given the set of observed labels ${\tilde{y}_{ij}}$). $P(\tilde{Y}_{ij}=k|Y_i=c)$ denotes the probability that worker $j$ annotates the item $i$ with label $k$ while the true label is $c$. They proposed to jointly estimate the distributions of $P$ and $Q$ by minimizing the entropy of the observed workers' labels conditioned on the true labels.

\begin{equation}
\min\limits_Q \max\limits_P H(\tilde{Y}|Y )
\label{eq:minmaxopt}
\end{equation}

The study solved this problem by converting this formulation into the dual form. They introduced two Lagrange multipliers $\sigma_j(c,k)$ and $\tau_i(c,k)$. Intuitively, $\sigma_j(c,k)$ is the measure of ability of worker $j$ and $\tau_i(c,k)$ is the intrinsic difficulty of item $i$. The variable $[\tau_i]$ is a confusion matrix of item $i$, where its $(c,k)-th$ entry measures how likely item $i$ in class $c$ is labeled as class $k$ by a randomly chosen worker \cite{Zhou_2015,Zhou_2014}. 

The minmax formulation by Zhou \etal \cite{Zhou_2014,Zhou_2015} provides a principled framework to estimate the difficulty of each sentence. We propose to use the difficulty measure $[\tau_i]$ to design the curriculum. The measure of difficulty $d_i$ is estimated with the ratio between the trace of $[\tau_i]$  and the sum of all the elements in the matrix.

\begin{equation}
d_i = \frac{\sum\limits_k\tau_i(k,k)}{\sum\limits_c\sum\limits_k\tau_i(c,k)}
\label{eq:minmaxdiff}
\end{equation}

The regularized minmax conditional entropy formulation proposed by Zhou \etal  \cite{Zhou_2014,Zhou_2015} finds  the item confusion matrix $[\tau_i]$ for both ordinal \cite{Zhou_2014} and categorical \cite{Zhou_2015} labels. After estimating the corresponding $[\tau_i]$, we find the difficulty metric using Equation \ref{eq:minmaxdiff} for regression, binary classification and multi-class classification problems.

\section{Experimental Evaluation}
\label{sec:experiment}
This section describes the experiments conducted to assess the performance of speech emotion recognition using the proposed curriculum learning schemes. This section introduces the database and the feature set used in the experiments (Sec. \ref{ssec:database}). The section also describes acoustic features (Sec. \ref{ssec:feat}) and classifier architecture used in the evaluation (Sec. \ref{ssec:classifier}). 

\subsection{The MSP-Podcast Database}
\label{ssec:database}

This study relies on the MSP-Podcast corpus collected at the University of Texas at Dallas\cite{Lotfian_201x}. The database includes a large set of speech segments from podcast recordings available in audio sharing websites. The podcasts are selected from various topics including politics, sports, talk shows, and movies, including a broad range of emotions. The podcasts are segmented into speech turns using a speaker diarization tool. We implement an automatic process that selects only speech segments with one speaker, without noise, background music, or phone quality audio. The duration of the segments are between 2.75s and 11s. Following the ideas described in Mariooryad \etal \cite{Mariooryad_2014_3}, we use different machine learning formulations to retrieve segments from the pool of available segments conveying target emotion behaviors. Then, these segments are annotated with emotional labels. The data collection process is an ongoing effort, where the current study uses version 1.0 of the corpus. This set includes 20,045 speech segments (34 hrs, 15 min). We have manually annotated the speaker identity of 244 speakers (16,026 segments). We use segments from 50 speakers as our test set (6,069 segments), and data from 15 speakers as our development set (2,226 segments). The train set includes the rest of the corpus (11,750 segments). This data partition attempts to create speaker independent datasets for train, test, and development sets. 

The speech segments are annotated with emotional labels using an improved version of the crowdsourcing method introduced by Burmania \etal \cite{Burmania_2016_2}. Within the perceptual evaluation, the raters are asked to choose emotional attributes (arousal, valence and dominance) using a seven-point Likert scale. Then, they select the primary emotion from anger, sadness, happiness, surprised, fear, disgust, contempt, and neutral. They can also choose other if none of the previous labels are suitable. The current version of the corpus is not balanced across emotional classes, with many happy and neutral sentences and very few fear sentences. Therefore, we remove the label \emph{fear} from the target emotion classes and consider it as if the raters had selected \emph{other} as the perceived emotion. While not used in this study, the annotation also includes secondary emotions where annotators choose all the emotions relevant to the speech segment. Each speech segment is annotated by at least five annotators.

\subsection{Features}
\label{ssec:feat}

All the emotion recognition problems are implemented with the feature set proposed for the \emph{computational paralinguistics challenge} (ComParE) at Interspeech 2013 \cite{Schuller_2013}. The feature set includes \emph{low-level descriptors} (LLDs) such as fundamental frequency and \emph{Mel-frequency cepstral coefficients} (MFCCs). These frame-based features are used to estimate statistics over a speech segment (e.g., mean of the fundamental frequency). The approach creates a 6,373D feature vector per speech segment, regardless of its length. The study by Schuller \etal \cite{Schuller_2013} provides more information about this feature set. The features are extracted with the open-source OpenSMILE toolkit \cite{Eyben_2010_2}.

\subsection{Implementation of Machine-Learning Frameworks}
\label{ssec:classifier}

The evaluation of the curriculum learning considers a DNN with the same architecture across the three machine-learning problems considered in this study (Sec. \ref{ssec:defproblem}). The only difference is the output layer, which is implemented according to the problem. The architecture is a fully connected feed forward neural network with two hidden layers, each of them implemented with 1,024 nodes. The activation function corresponds to \emph{rectified linear unit} (ReLU). The input corresponds to the 6,373D segment-label feature vector described in Section \ref{ssec:feat}. The output layer is added on top of the second hidden layer. The output layer is the identity activation function for the regression problems, and the softmax layer for binary and multi-class classification tasks. The regression network is trained by minimizing the \emph{mean square error} (MSE). The binary and multi-class networks are trained to minimize the cross-entropy cost function. The softmax layer gives a vector with a score for each emotional class. The evaluation uses Keras with TensorFlow as backend to implement and train the models. We rely on \emph{adaptive moment estimation} (ADAM) \cite{Kingma_2014_2} for the optimization of the parameters of the network.

Based on the measure of difficulty, we divide the training set into five bins, where the first bin contains the easiest samples. The training process starts with the easiest bin and continue by adding more difficult bins to the training set. After adding each bin to the training set, the network is trained for 50 iterations. We find the optimal learning rate for each bin by maximizing the performance on the development set. In our search, we consider the following values for the learning rates: 0.1, 0.05, 0.01, 0.005, 0.001, 0.0005, 0.0001, 0.00005, 0.00001, 0.000005, and 0.000001. We start with bin 1,  finding the optimal learning rate by considering only the performance using the training samples on bin 1. The learning rate for bin 1 does not change during the rest of the search. Next, we add the training samples of bin 2, finding the optimal learning rate for bin 2. This process is repeated until we find the optimal learning rates for all the five bins. Although this search does not guarantee an optimal solution, we choose this method since it can be efficiently implemented. The optimal combination of learning rates with this approach monotonically decrease from the first bin to the last training set (all bins): 0.001 (bin 1), 0.0005 (bin 2), 0.0001 (bin 3), 0.00005 (bin 4), and 0.00001 (bin 5).

We consider two baseline frameworks to evaluate the proposed curriculum learning approaches. The first approach does not consider any curriculum, training the models with all the data. This framework is implemented with 100 iterations with a learning rate of $0.0005$. This value was also selected based on the results on the development set. The second baseline uses a curriculum where the bins are created at random. This baseline is implemented using the same learning rates used for the models trained with the proposed curriculum policies.

\section{Results}
\label{sec:perfclass}
This section describes the results obtained on the three machine-learning problems for emotion recognition (regression, binary classification, and multi-class classification). We train the DNNs 10 times, using different random initializations. We report the average results, evaluating the performance metric using the one-tail, population mean t-test over the 10 trials. We assert statistical significance at $p$-value$\leq$ 0.05. We denote with an asterisk ($^*$) when a model trained with curriculum learning is statistically better than the baseline model trained without a curriculum, and with a circle ($^\circ$) when a model trained with curriculum learning is statistically better than the baseline trained with a random curriculum (see Tables \ref{tab:reggtab}, \ref{tab:binarytab} and \ref{tab:categtab}).

\subsection{Regression of Emotional Attributes}
\label{ssec:regression}

The first experimental evaluation demonstrates the role of curriculum learning on predicting emotional attributes with regression models. The performance is computed with \emph{concordance correlation coefficient} (CCC) for arousal, valence and dominance. The CCC is a metric of  agreement between two interval variables by considering the correlation and difference between them. We employ this metric to compare the true and predicted emotional attribute values. This metric has been previously used in related studies, including the AVEC 2016 depression, mood and emotion recognition challenge \cite{Valstar_2016}

\begin{table}[t]
\centering
\caption{Results of regression models for arousal (\emph{Aro}), valence ( \emph{Val}), and dominance (\emph{Dom}).The asterisk [$^*$] and circle [$^\circ$] indicate the approach outperforms the baselines \emph{w/o curriculum}  and \emph{with random curriculum}, respectively. We assert significance at $p$-value$\leq$ 0.05.}
\begin{tabular*}{1\columnwidth}{@{\extracolsep{\fill}} l @{\hspace{0.1cm}}| l @{\hspace{0.2cm}} l @{\hspace{0.2cm}} l}
\hline
& Aro.  & Val. & Dom.\\
& [CCC]  & [CCC] & [CCC]\\
\hline\hline
w/o curriculum &0.719&0.293& 0.684\\
With random curriculum & 0.721 & 0.290 & 0.683 \\
\hline
Criterion 1-Error of predicted label& 0.723&0.298&0.684\\
Criterion 2-Disagreement between annotators& 0.728$^{*}$& 0.304$^{\circ}$ & 0.692$^{\circ}$\\
Criterion 3-Minmax entropy& 0.735$^{*\circ}$&0.313$^{*\circ}$&0.696$^{*\circ}$\\
\hline
\end{tabular*}
\label{tab:reggtab}
\end{table}

Table \ref{tab:reggtab} shows the average CCC values for arousal, valence and dominance across the 10 trials. The results show that the condition with the highest CCC values corresponds to the regression models trained with curriculum learning using the criterion 3 (i.e., minmax entropy). The results are statistically better than both baseline methods. The table also shows that randomly selecting the training bins does not lead to significant improvements. Criterion 1 is less effective than criteria 2 and 3. These results demonstrate the importance of quantifying the disagreement between evaluators to assess the difficulty of the samples. We can effectively achieve this goal by considering individual evaluations assigned to the samples. 

\begin{figure*}[tb]
\centering
\subfigure[Arousal]{
   \includegraphics[width=5.7cm,]{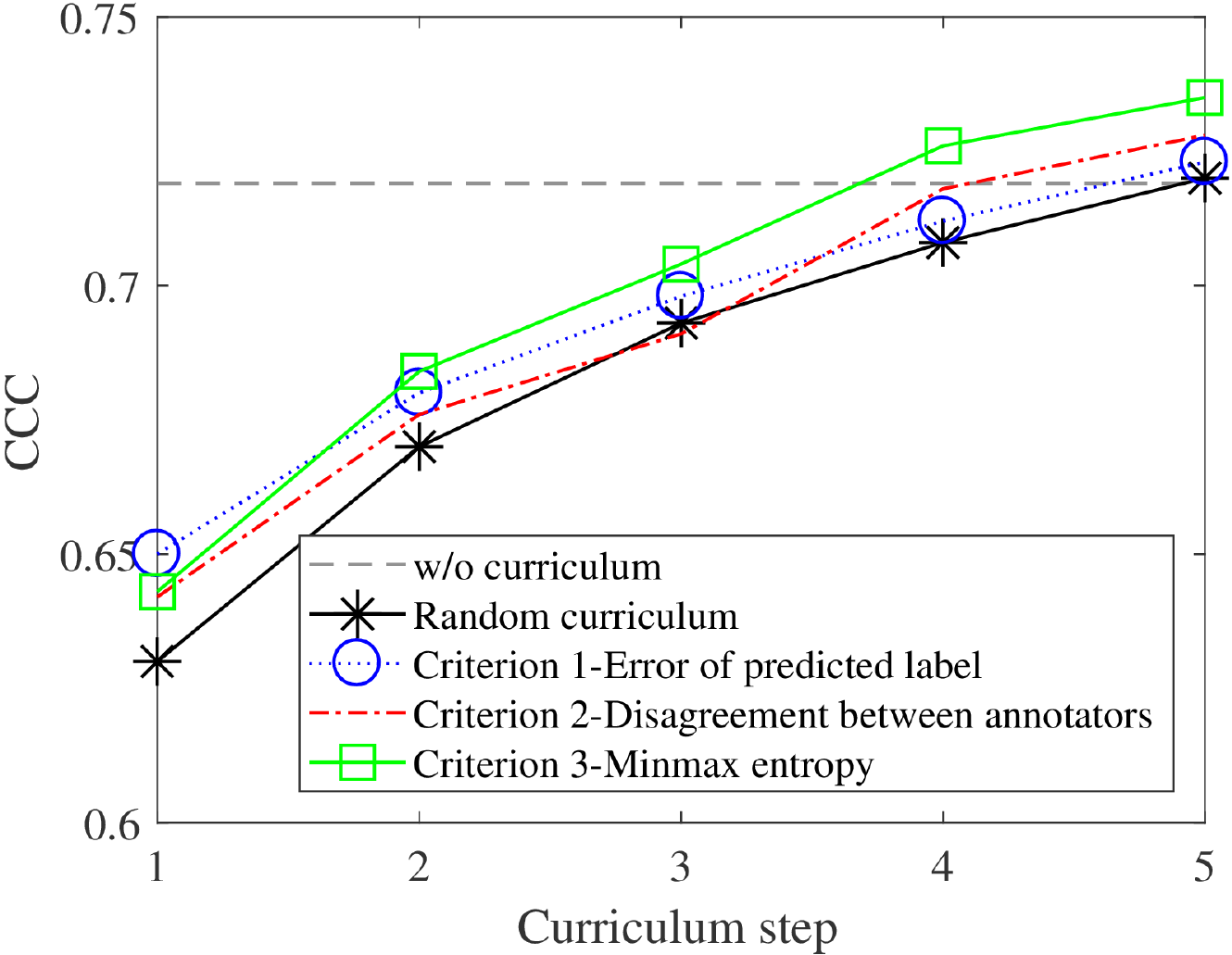}
   \label{fig:Reg_aro}
}
\hfill
\subfigure[Valence]{
   \includegraphics[width=5.7cm,]{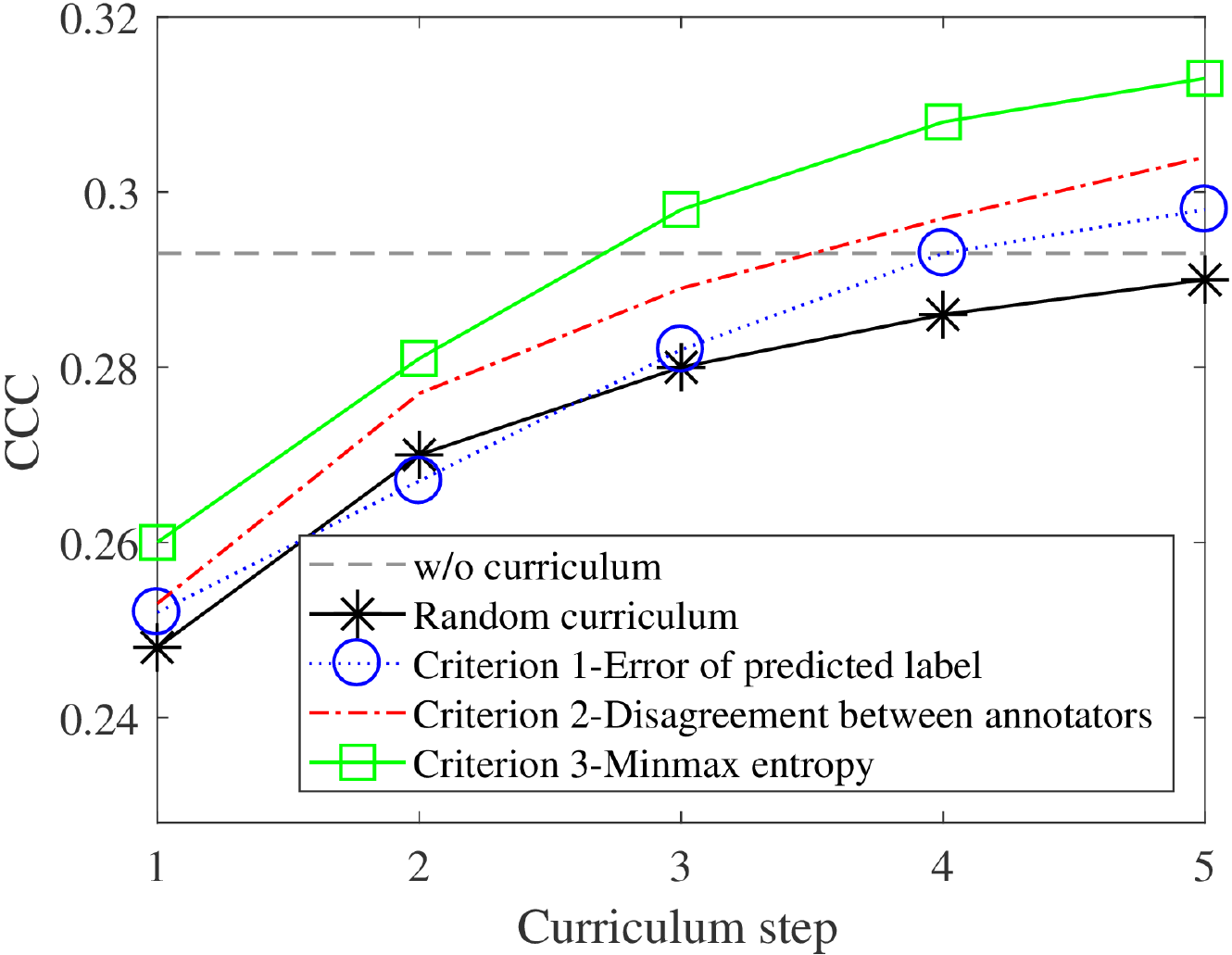}
   \label{fig:Reg_val}
}
\hfill
\subfigure[Dominance]{
   \includegraphics[width=5.7cm,]{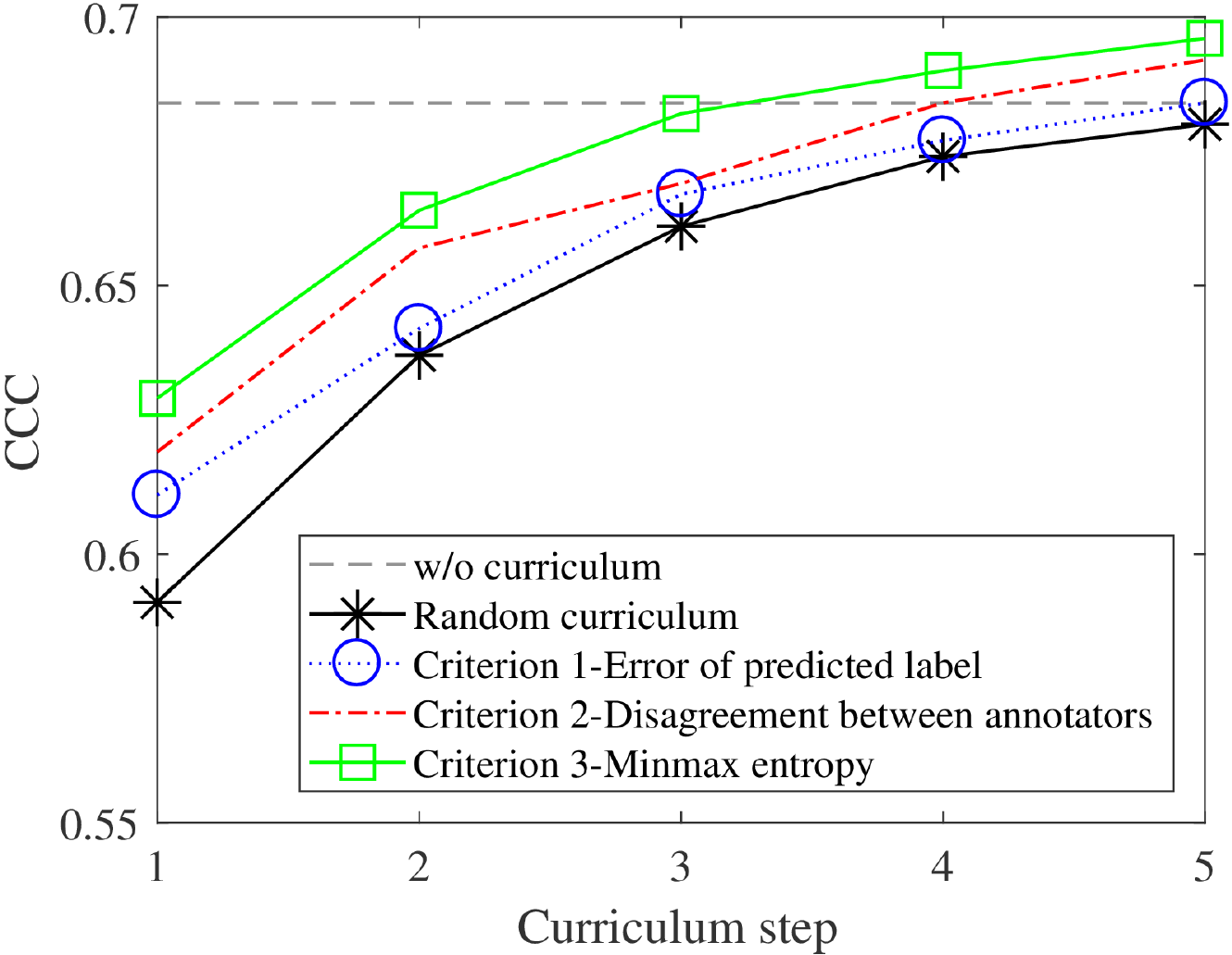}
   \label{fig:Reg_dom}
}
\caption{Intermediate results for regression models. The results are obtained on the test set at different steps of the training process using curriculum learning. More difficult samples are added at each step. The dashed lines indicate the performance of regression models trained without curriculum learning.}
\label{fig:learncurve_reg}
\end{figure*}

Figure \ref{fig:learncurve_reg} shows the CCC values for the regression models for emotional attributes by following different curriculum policies. The figure shows that the performance increases as we effectively add more difficult samples in the training set. The criterion 3 based on the minmax entropy framework achieves the highest performance for arousal, valence, and dominance. The performance for this approach is consistently better than a system trained without curriculum learning.

\subsection{Binary Classification of Emotional Attributes}
\label{ssec:binary}

We also evaluate the role of curriculum learning on binary classification of emotional attributes (e.g., low arousal versus high arousal). The test set is divided into two classes for high and low values of a given attribute based on the median split obtained on the training set. This method makes the test set almost balanced.

\begin{table}[tbp!]
\centering
\caption{Results of binary classification for arousal (\emph{Aro}), valence ( \emph{Val}), and dominance (\emph{Dom}).The asterisk [$^*$] and circle [$^\circ$] indicate the approach outperforms the baselines \emph{w/o curriculum}  and \emph{with random curriculum}, respectively. We assert significance at $p$-value$\leq$ 0.05.}
\begin{tabular*}{1\columnwidth}{@{\extracolsep{\fill}} l @{\hspace{0.01cm}}| l @{\hspace{0.1cm}}l @{\hspace{0.1cm}}l}
\hline
& Aro.  &Val.&Dom.\\
& [F-score] &[F-score]&[F-score]\\
\hline\hline
w/o curriculum & 0.779&0.592& 0.684\\
With random curriculum & 0.771 & 0.590& 0.685 \\
\hline
Criterion 1-Error of predicted label& 0.783&0.607$^{*\circ}$&0.684\\
Criterion 2-Disagreement between annotators&0.789$^{*\circ}$&0.615$^{*\circ}$ &0.692$^*$\\
Criterion 3-Minmax entropy &0.791$^{*\circ}$&0.616$^{*\circ}$&0.696$^{*\circ}$\\

\hline
\end{tabular*}
\label{tab:binarytab}
\end{table}

Table \ref{tab:binarytab} lists the average F-score for arousal, valence, and dominance across the 10 trails. The table demonstrates that using curriculum learning with criteria 2 and 3 achieves statistically significant improvements over a model trained without curriculum learning or with randomly selected bins. However, the best performance is also obtained with criterion 3, which uses the policy based on the minmax entropy framework.  Criterion 1 is only effective for valence, showing that quantifying the difficulty of the samples based on pre-trained models is not the best approach for speech emotion recognition. 

\begin{figure*}[tb]
\centering
\subfigure[Arousal]{
   \includegraphics[width=5.7cm,]{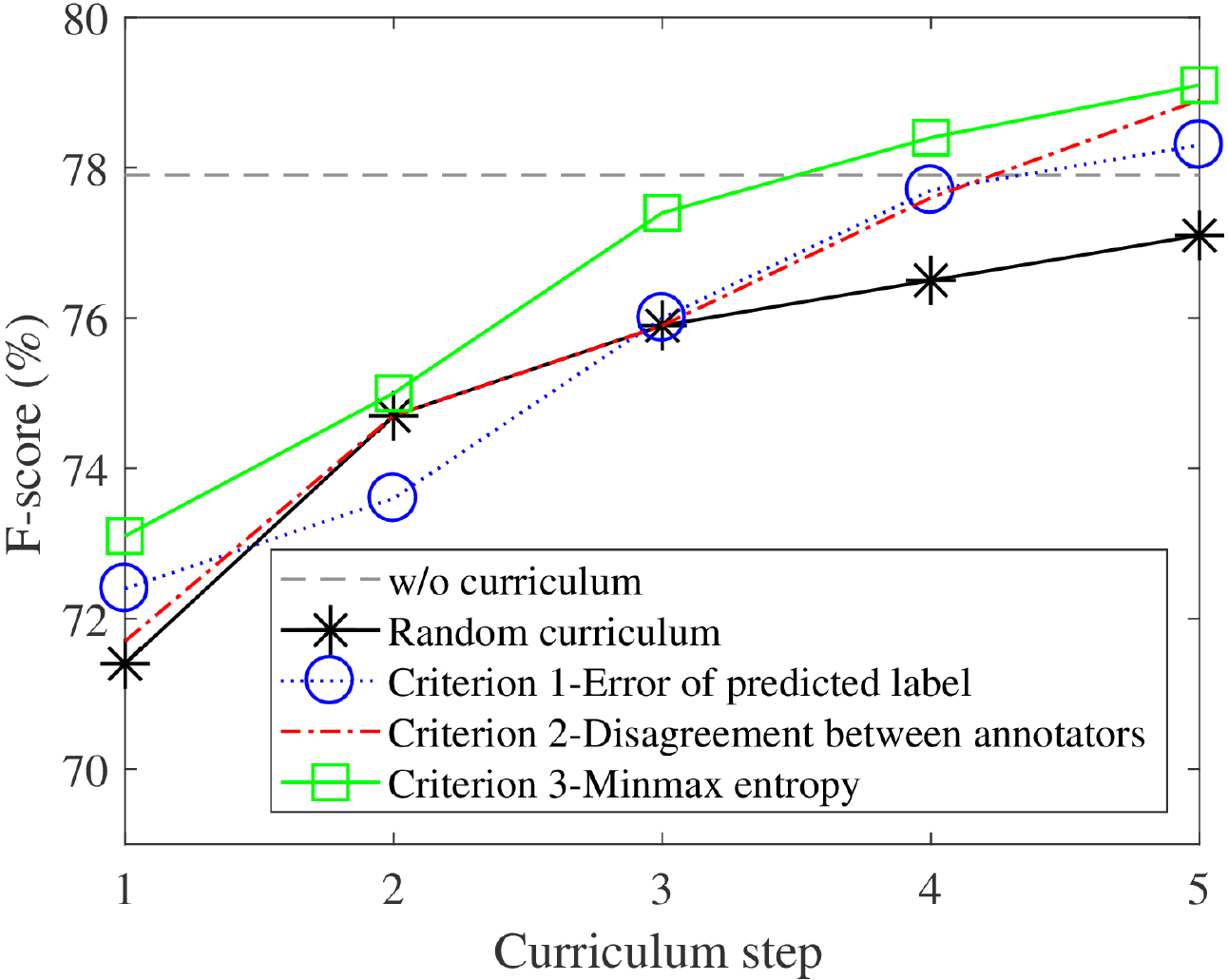}
   \label{fig:bin_aro}
}
\hfill
\subfigure[Valence]{
   \includegraphics[width=5.7cm,]{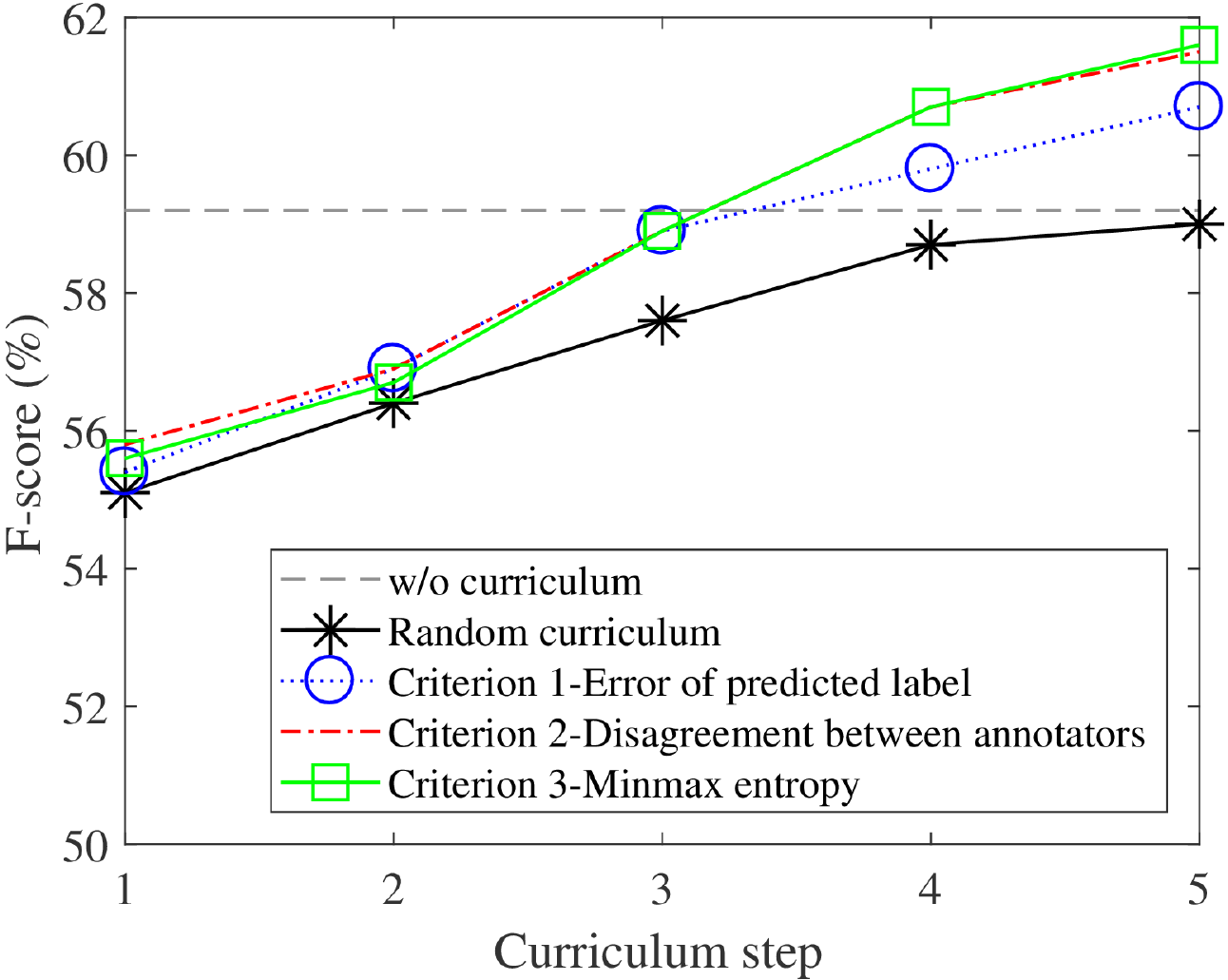}
   \label{fig:bin_val}
}
\hfill
\subfigure[Dominance]{
   \includegraphics[width=5.7cm,]{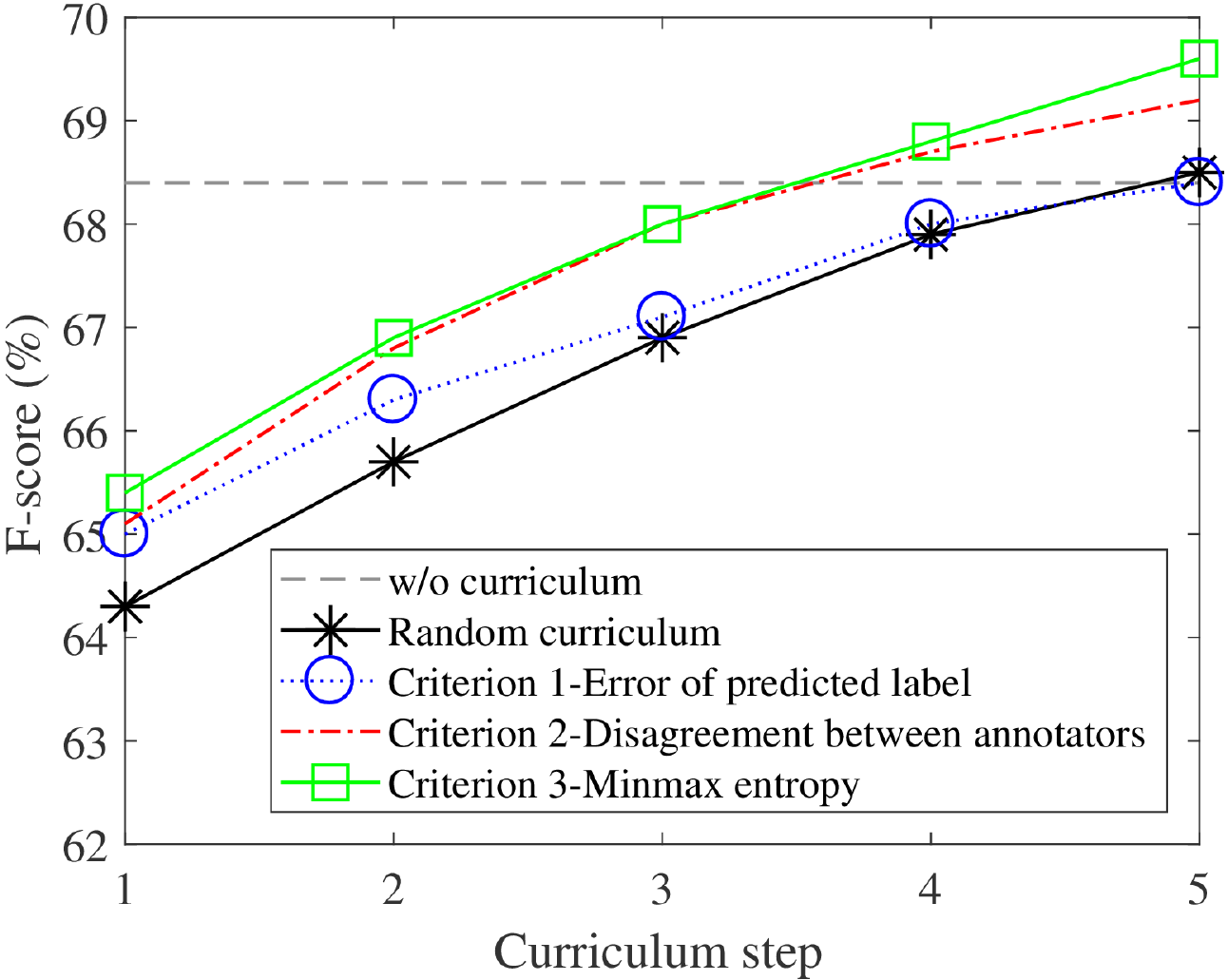}
   \label{fig:bin_dom}
}
\caption{Intermediate results for binary classification. The results are obtained on the test set at different steps of the training process using curriculum learning. More difficult samples are added at each step. The dashed lines indicate the performance of binary classifiers trained without curriculum learning.}
\label{fig:learncurve_binary}
\end{figure*}

Figure \ref{fig:learncurve_binary} shows the F-score values for attribute based binary classification tasks following the three different curriculum policies. It also shows the performance of a model trained without a curriculum and with a curriculum using randomly selected bins. Similar to the results with regression problems, criterion 3 achieves the best performance for arousal, valence, and dominance, obtaining important improvements over the baselines. 

\subsection{Multi-class Categorical Emotion Classification}
\label{ssec:Multiclass}

The third problem involves the classification of categorical emotions. 
The evaluation consider the following five classes: happiness, anger, sadness, disgust, and neutral state. The ground truth labels for the test set are generated by finding the majority vote between all the annotators. Samples that do not reach agreement with this rule are discarded from this evaluation.

\begin{table}[tbp!]
\centering
\caption{Performance of categorical emotion classification. The asterisk [$^*$] and circle [$^\circ$] indicate the approach outperforms the baselines \emph{w/o curriculum}  and \emph{with random curriculum}, respectively. We assert significance at $p$-value$\leq$ 0.05.}
\begin{tabular*}{1\columnwidth}{@{\extracolsep{\fill}} l | c c c}
\hline
& F-score [\%]\\
\hline\hline
w/o curriculum & 39.7\\
With random curriculum & 39.6 \\
\hline
Criterion 1-Error of predicted label& 40.8\\
Criterion 2-Disagreement between annotators& 41.4$^{*\circ}$\\
Criterion 3-Minmax entropy& 42.1$^{*\circ}$\\
\hline
\end{tabular*}
\label{tab:categtab}
\end{table}

Table \ref{tab:categtab} shows the average F-score of the five class categorical emotion classification across the 10 trails. The results are consistent with the findings presented for regression and binary classification problems. Curriculum learning using policies that quantify the agreement between evaluators (criteria 2 and 3) provides statistically significant improvement over the baseline methods. For criterion 3, the use of curriculum learning leads to an absolute improvement of 2.4\% over a model trained without curriculum learning. This gain corresponds to a 6\% relative improvement over the baseline, which is obtained by just changing the order in which the samples are introduced during the training process.

\begin{figure}[tbp]
\centering
 \includegraphics[width=2.5in]{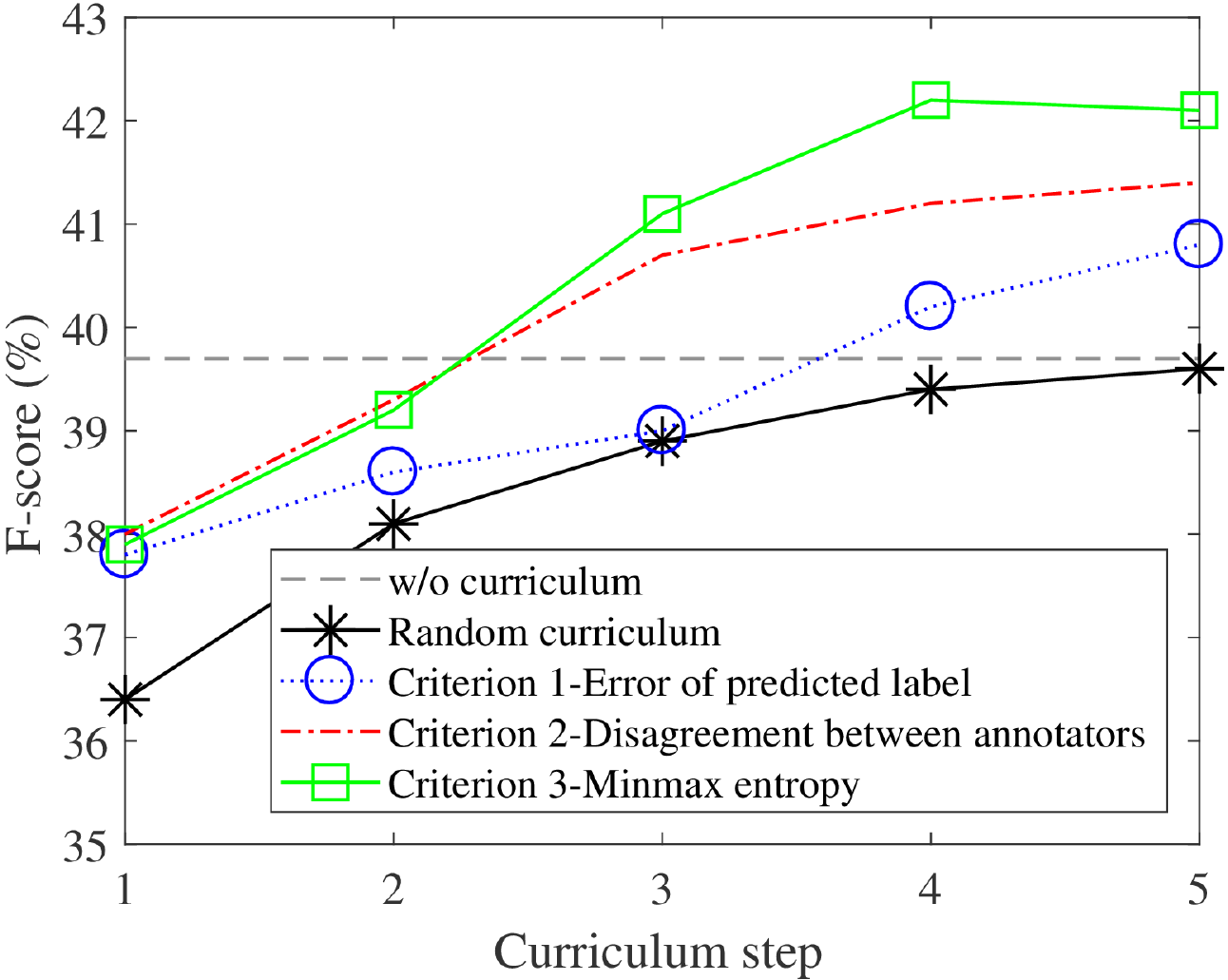}
 \caption{Intermediate results for categorical emotions. The results are obtained on the test set at different steps of the training process using curriculum learning. More difficult samples are added at each step. The dashed line indicates the performance of a classifier trained without curriculum learning.}
\label{fig:learncurve}
\end{figure}

Figure \ref{fig:learncurve} shows the F-score curve for the classification of categorical emotions by following different curriculum policies. It is clear that criterion 3 based on the minmax entropy framework also achieves the best performance for this task. The use of criteria 2 and 3 leads to statistically significant improvements over the baselines. We also observe that for criterion 3, the last bin does not improve the performance, suggesting that the labels in this bin are noisy and less informative.

\begin{figure*}[tbh]
\centering
\includegraphics[width=0.45\columnwidth]{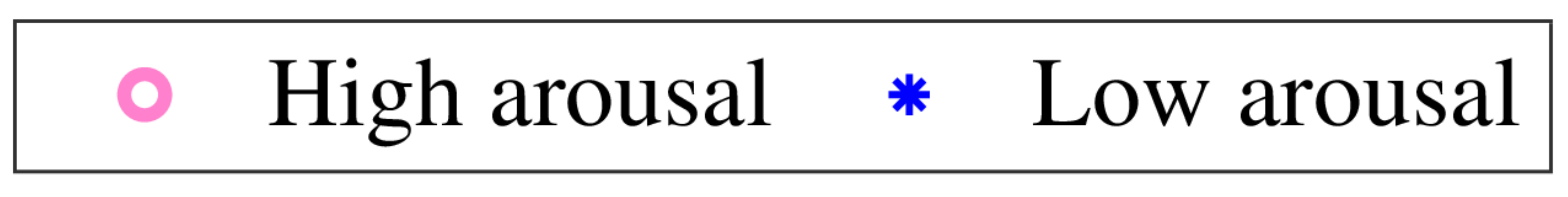}\\
\subfigure[Arousal - Criterion 1 - bin 1]
{
  \includegraphics[width=0.5\columnwidth]{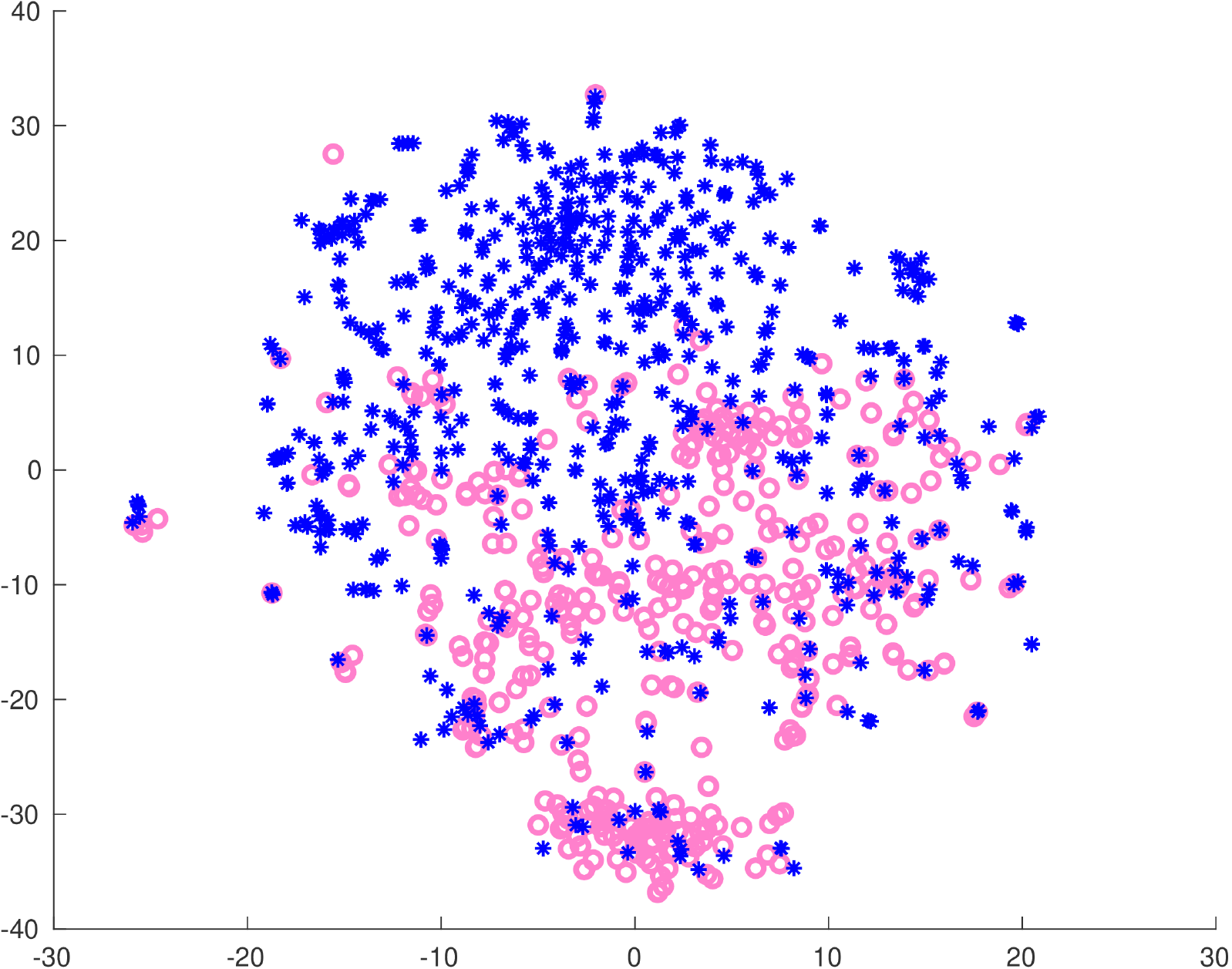}
  \label{fig:easy_EOP_act}
}
\subfigure[Arousal - Criterion 1 - bin 3]
{
  \includegraphics[width=0.5\columnwidth]{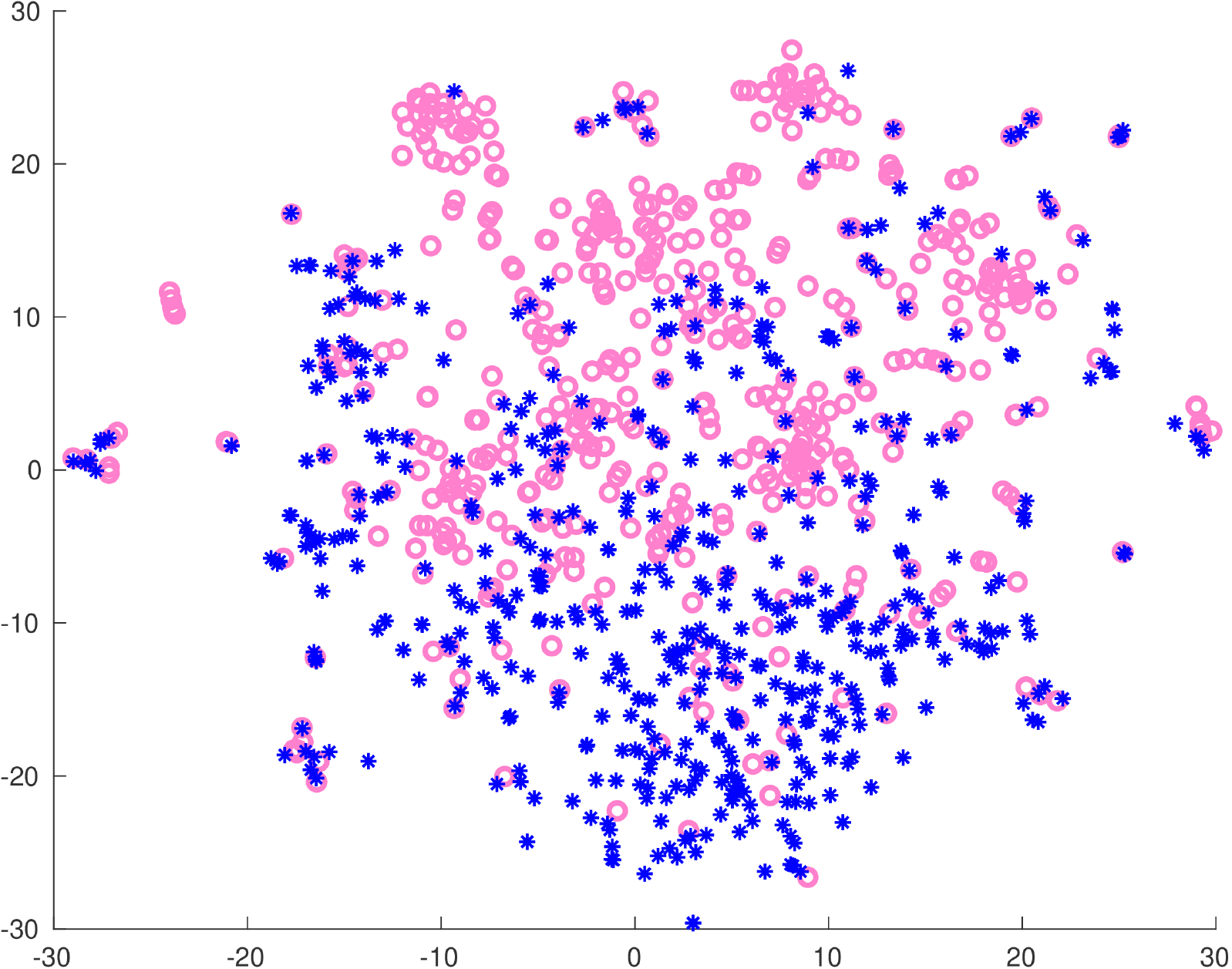}
  \label{fig:med_EOP_act}
}
\subfigure[Arousal - Criterion 1 - bin 5]
{
  \includegraphics[width=0.5\columnwidth]{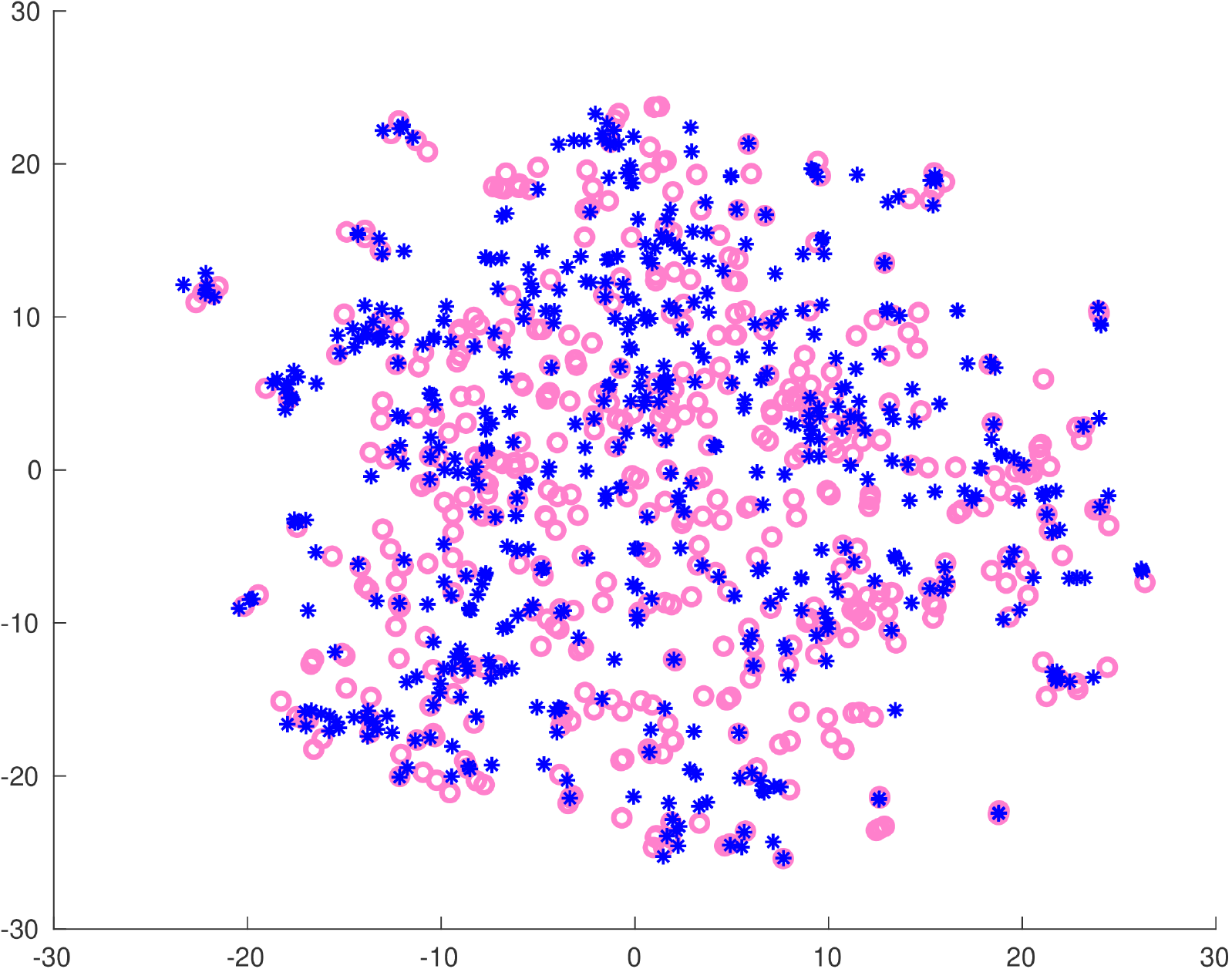}
  \label{fig:hard_EOP_act}
}
\subfigure[Arousal - Criterion 3 - bin 1]
{
  \includegraphics[width=0.5\columnwidth]{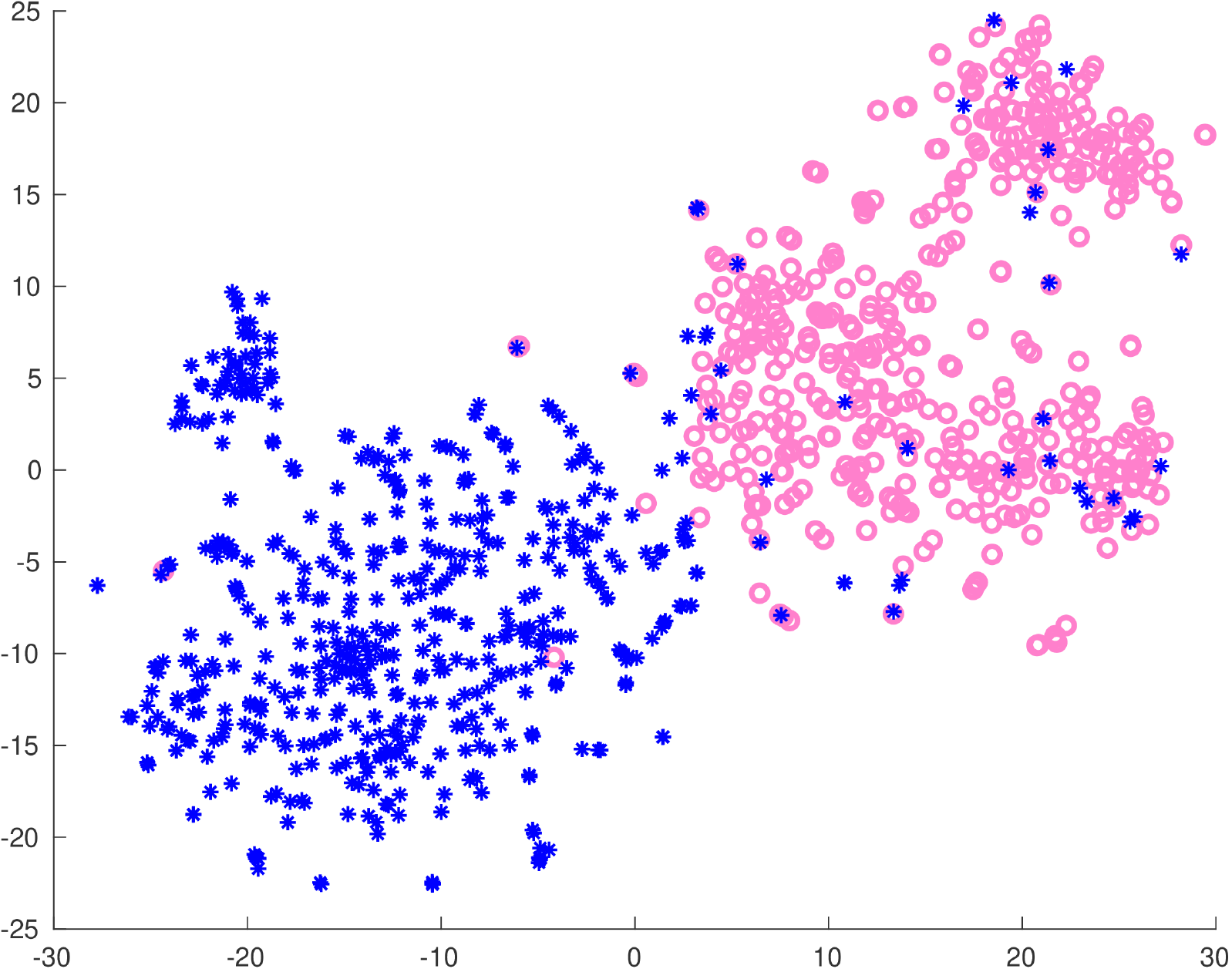}
  \label{fig:easy_ME_act}
}
\subfigure[Arousal - Criterion 3 - bin 3]
{
  \includegraphics[width=0.5\columnwidth]{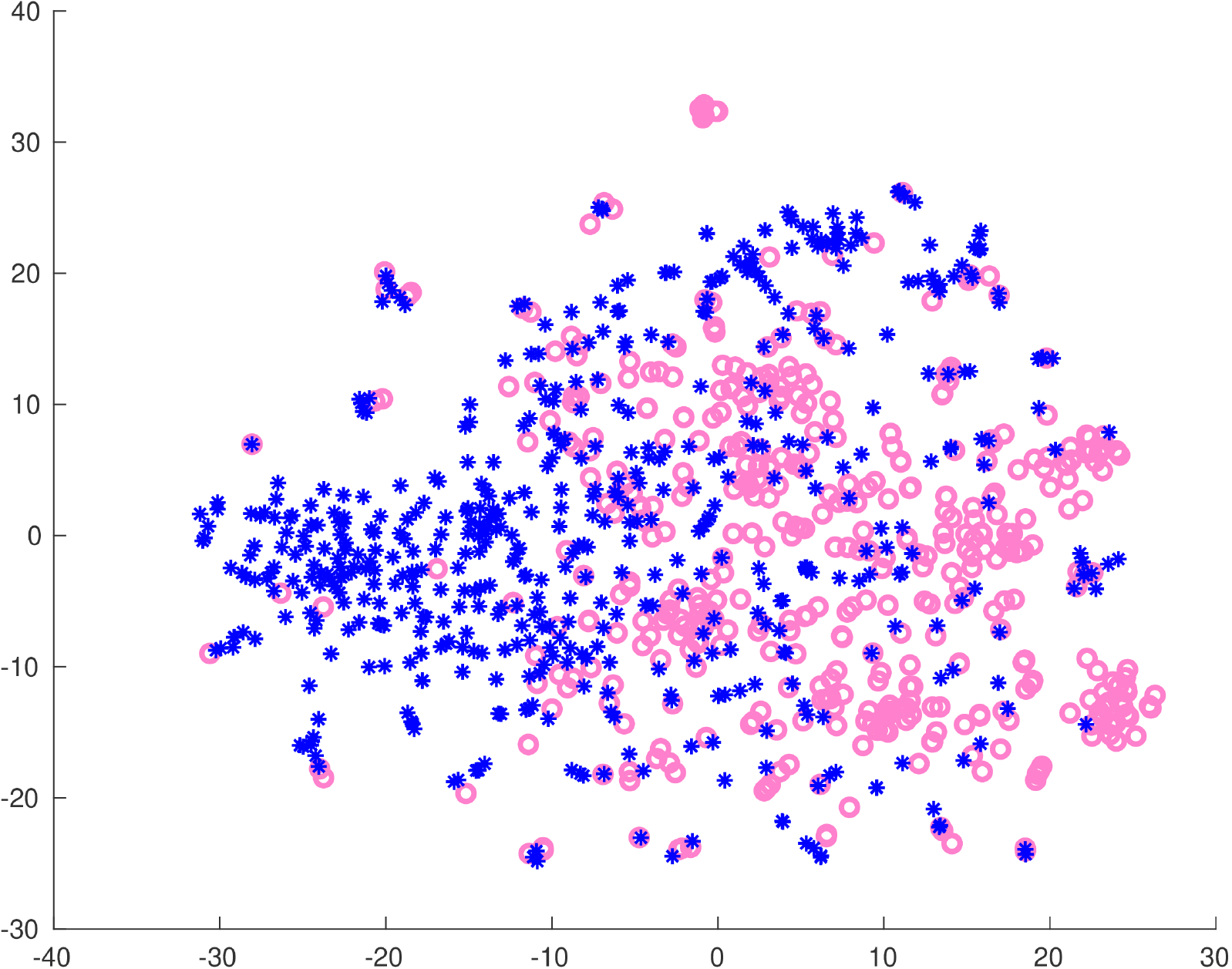}
  \label{fig:med_ME_act}
}
\subfigure[Arousal - Criterion 3 - bin 5]
{
  \includegraphics[width=0.5\columnwidth]{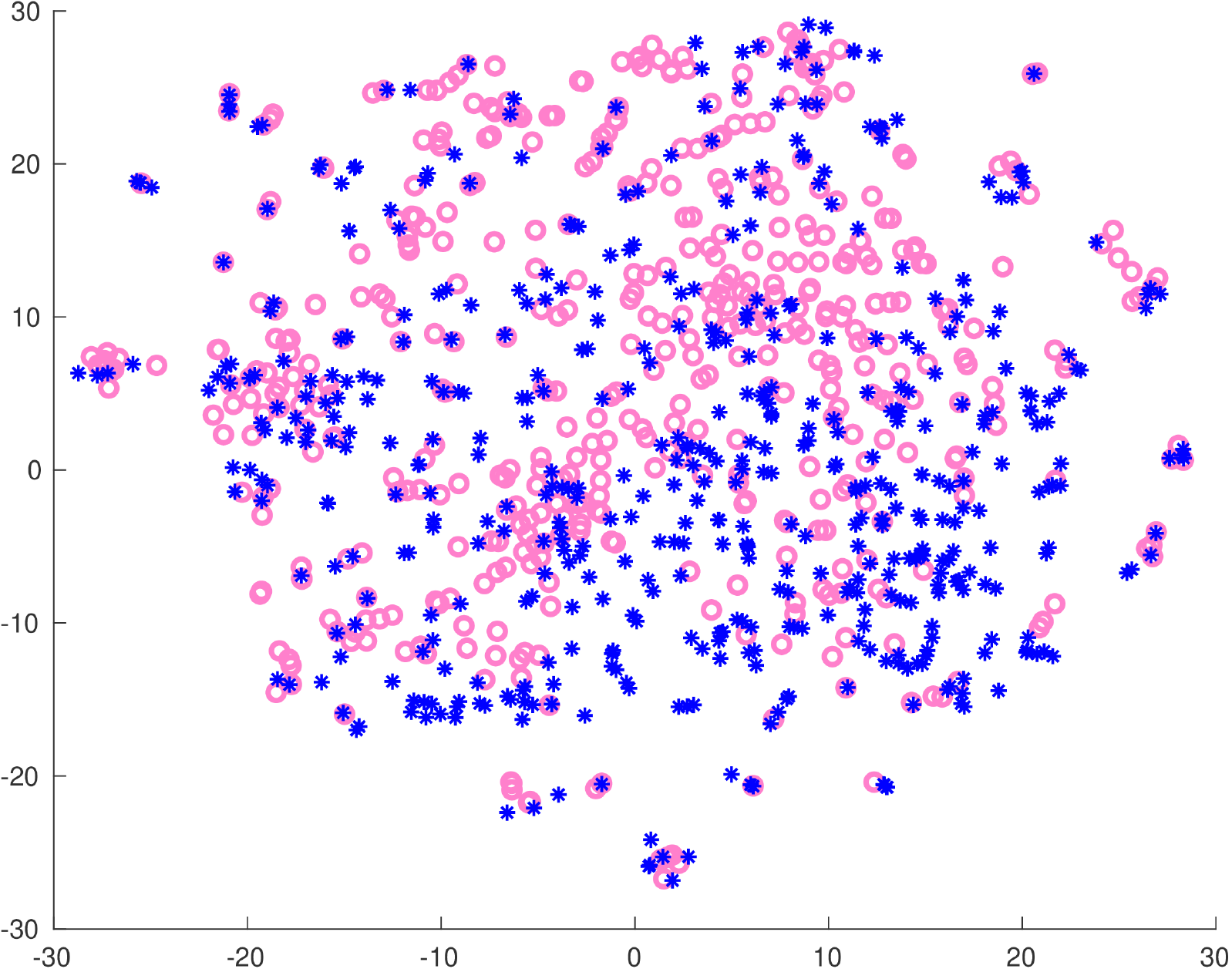}
  \label{fig:hard_ME_act}
}
\caption{Visualization using the t-SNE toolkit of acoustic features for arousal in binary classification tasks for bin 1 (easy), bin 2 (medium difficulty), and bin 3 (hard). The top figures correspond to criterion 1 (i.e., error of predicted labels) and the bottom figures correspond to criterion 3 (i.e., minmax entropy).}
\label{fig:tsnecat_act}
\end{figure*}


\subsection{Analysis of Feature Representation}
\label{ssec:featanalys}

This section visualizes how the difficulty measures used to build the proposed curriculum is reflected on the features. This analysis is applicable to classification problems, where we only consider binary classification of emotional attributes. The analysis relies on the t-SNE method proposed by Van Der Maaten and Hinton \cite{Maaten_2008}, which is a useful tool to visualize high dimensional data. This toolkit is used to reduce the dimension of the features from 6,373 to 2. We expect that the difficulty measure used to build the curriculum is reflected in the feature representation, where the different classes in the first bin (i.e., the easiest samples) are better separated than the classes in other bins (i.e., more difficult samples). This feature visualization provides another venue to evaluate the proposed criteria for curriculum learning. The analysis considers criterion 1 (error of predicted error) and criterion 3 (minmax entropy), which provided the worse and best policies for curriculum learning in previous sections.

We report the results for arousal in the binary classification problem. Figure \ref{fig:tsnecat_act} shows the feature representation for arousal, where each point is associated with one sentence. The color of the points represent the category of the sentence (low arousal versus high arousal). Figures \ref{fig:easy_EOP_act}, \ref{fig:med_EOP_act}, and \ref{fig:hard_EOP_act} show the results when using criterion 1 for bin 1 (easy), bin 3 (medium difficulty) and bin 5 (hard), respectively. The classes for the easy samples have a clear separation. Figure \ref{fig:hard_EOP_act} shows that the overlap between classes increases in bin 5, which has the most difficult examples.  The correlation between task difficulty and ambiguity in the feature domain was expected since the difficulty is derived from the prediction of pre-trained classifiers, which directly rely on the discrimination of the features. This connection is not presupposed for the minmax entropy framework (criterion 3), which relies on the disagreement between evaluators. Figures \ref{fig:easy_ME_act}, \ref{fig:med_ME_act}, and \ref{fig:hard_ME_act} show the results using criterion 3 for bin 1 (easy), bin 3 (medium difficulty) and bin 5 (hard), respectively. It is interesting to observe similar trends, even when the difficulty metric is not derived from pre-trained classifiers. 


\section{Conclusions}
\label{sec:conclusion}
This study proposed the use of curriculum learning for speech emotion recognition. Since the policies to determine the difficulty of emotional sentences are not straightforward as in other problems, the study explored different methods to design the curriculum. The study proposed to quantify the difficulty level of the sentences by relying on results from pre-trained models on the training set, or by considering inter-evaluator agreement metrics under the assumption that ambiguous sentences for human are also ambiguous for machines. The experimental evaluation considered three formulations of speech emotion recognition: regression of emotional attributes, binary classification of emotional attributes, and classification of emotional categories. The results demonstrated the benefits of using curriculum learning in speech emotion recognition, showing consistent improvements over baselines trained with randomly selected bins or without curriculum learning. The most successful policy for building the curriculum considers the agreement level of the annotations by estimating the expertise of the labelers. This approach relies on the minmax entropy framework that learns a latent variable describing the difficulty of the speech samples evaluated by the raters. The machine-learning models trained with this curriculum achieved significant improvements over the baseline methods. 

As a future direction of the curriculum learning framework proposed in this study, we will use the difficulty measure to find training examples that negatively affect the performance of the models. Abdelwahab and Busso \cite{Abdelwahab_2017_2} showed that selecting a subset of the data for supervised adaptation of speech emotional models led to improvements over results obtained when the entire adaptation set was used. By removing these samples, we expect to increase the classification performance, since these examples can be too difficult to learn due to unreliable or incorrect labels. Likewise, there are important parameters that we have not investigated, including the optimum number of difficulty bins, and the optimum number of epochs. Adjusting these parameters may lead to further improvements in classification performance. Finally, we will explore whether curriculum learning is still effective as the size of the training set increases. The collection of the MSP-Podcast is an ongoing effort in our laboratory, which will allow us to evaluate the approach in the future with a larger training set. 

\vspace{-0.2cm}


%

%
%

\ifCLASSOPTIONcompsoc
  \section*{Acknowledgments}
\else
  \section*{Acknowledgment}
\fi

This study was funded by the National Science Foundation (NSF) CAREER grant IIS-1453781.

\ifCLASSOPTIONcaptionsoff
  \newpage
\fi

\bibliographystyle{IEEEtran}
\bibliography{reference}

\begin{thebibliography}{10}
\providecommand{\url}[1]{#1}
\csname url@samestyle\endcsname
\providecommand{\newblock}{\relax}
\providecommand{\bibinfo}[2]{#2}
\providecommand{\BIBentrySTDinterwordspacing}{\spaceskip=0pt\relax}
\providecommand{\BIBentryALTinterwordstretchfactor}{4}
\providecommand{\BIBentryALTinterwordspacing}{\spaceskip=\fontdimen2\font plus
\BIBentryALTinterwordstretchfactor\fontdimen3\font minus
  \fontdimen4\font\relax}
\providecommand{\BIBforeignlanguage}[2]{{%
\expandafter\ifx\csname l@#1\endcsname\relax
\typeout{** WARNING: IEEEtran.bst: No hyphenation pattern has been}%
\typeout{** loaded for the language `#1'. Using the pattern for}%
\typeout{** the default language instead.}%
\else
\language=\csname l@#1\endcsname
\fi
#2}}
\providecommand{\BIBdecl}{\relax}
\BIBdecl

\bibitem{Bartlett_2003}
M.~S. Bartlett, G.~Littlewort, I.~Fasel, and J.~R. Movellan, ``Real time face
  detection and facial expression recognition: Development and applications to
  human computer interaction,'' in \emph{Conference on Computer Vision and
  Pattern Recognition Workshop (CVPRW 2003)}, Madison, WI, USA, June 2003, pp.
  1--6.

\bibitem{Szwoch_2015}
M.~Szwoch and W.~Szwoch, ``Emotion recognition for affect aware video games,''
  in \emph{Image Processing {\&} Communications Challenges 6}, ser. Advances in
  Intelligent Systems and Computing, R.~Chora{\'{s}}, Ed.\hskip 1em plus 0.5em
  minus 0.4em\relax Cham: Springer International Publishing, 2015, vol. 313,
  pp. 227--236.

\bibitem{Obaid_2008}
M.~Obaid, C.~Han, and M.~Billinghurst, ````{Feed} the fish'': an affect-aware
  game,'' in \emph{Australasian Conference on Interactive Entertainment},
  Brisbane, Australia, December 2008.

\bibitem{Litman_2004}
D.~Litman and K.~Forbes-Riley, ``Predicting student emotions in computer-human
  tutoring dialogues,'' in \emph{ACM Association for Computational Linguistics
  (ACL 2004)}, Barcelona, Spain, July 2004, pp. 1--8.

\bibitem{Litman_2006}
D.~Litman and K.~{Forbes-Riley}, ``Recognizing student emotions and attitudes
  on the basis of utterances in spoken tutoring dialogues with both human and
  computer tutors,'' \emph{Speech communication}, vol.~48, no.~5, pp. 559--590,
  May 2006.

\bibitem{Abirached_2012}
B.~Abirached, Y.~Zhang, and J.-H. Park, ``Understanding user needs for serious
  games for teaching children with autism spectrum disorders emotions,'' in
  \emph{World Conference on Educational Media and Technology (EdMedia 2012)},
  Denver, CO, USA, June 2012, pp. 1054--1063.

\bibitem{Busso_2013}
C.~Busso, M.~Bulut, and S.~Narayanan, ``Toward effective automatic recognition
  systems of emotion in speech,'' in \emph{Social emotions in nature and
  artifact: emotions in human and human-computer interaction}, J.~Gratch and
  S.~Marsella, Eds.\hskip 1em plus 0.5em minus 0.4em\relax New York, NY, USA:
  Oxford University Press, November 2013, pp. 110--127.

\bibitem{Schuller_2010_2}
B.~Schuller, B.~Vlasenko, F.~Eyben, M.~W\"{o}llmer, A.~Stuhlsatz, A.~Wendemuth,
  and G.~Rigoll, ``Cross-corpus acoustic emotion recognition: Variances and
  strategies,'' \emph{IEEE Transactions on Affective Computing}, vol.~1, no.~2,
  pp. 119--131, July-Dec 2010.

\bibitem{Zhang_2011_2}
Z.~Zhang, F.~Weninger, M.~Wollmer, and B.~Schuller, ``Unsupervised learning in
  cross-corpus acoustic emotion recognition,'' in \emph{IEEE Workshop on
  Automatic Speech Recognition and Understanding (ASRU 2011)}, Waikoloa, HI,
  USA, December 2011, pp. 523--528.

\bibitem{Vogt_2006}
T.~Vogt and E.~Andr{\'e}, ``Improving automatic emotion recognition from speech
  via gender differentiation,'' in \emph{International conference on Language
  Resources and Evaluation (LREC 2006)}, Genoa,Italy, May 2006, pp. 1123--1126.

\bibitem{Schuller_2005}
B.~Schuller, R.~M\"uller, M.~Lang, and G.~Rigoll, ``Speaker independent emotion
  recognition by early fusion of acoustic and linguistic features within
  ensembles,'' in \emph{9th European Conference on Speech Communication and
  Technology (Interspeech 2005 - Eurospeech)}, Lisbon, Portugal, September
  2005, pp. 805--808.

\bibitem{Busso_2013_2}
C.~Busso, S.~Mariooryad, A.~Metallinou, and S.~Narayanan, ``Iterative feature
  normalization scheme for automatic emotion detection from speech,''
  \emph{IEEE Transactions on Affective Computing}, vol.~4, no.~4, pp. 386--397,
  October-December 2013.

\bibitem{Scherer_2003}
K.~Scherer, ``Vocal communication of emotion: A review of research paradigms,''
  \emph{Speech Communication}, vol.~40, no. 1-2, pp. 227--256, April 2003.

\bibitem{Mower_2009}
E.~Mower, A.~Metallinou, C.-C. Lee, A.~Kazemzadeh, C.~Busso, S.~Lee, and
  S.~Narayanan, ``Interpreting ambiguous emotional expressions,'' in
  \emph{International Conference on Affective Computing and Intelligent
  Interaction (ACII 2009)}, Amsterdam, The Netherlands, September 2009, pp.
  1--8.

\bibitem{Elman_1993}
J.~Elman, ``Learning and development in neural networks: The importance of
  starting small,'' \emph{Cognition}, vol.~48, no.~1, pp. 71--99, July 1993.

\bibitem{Turian_2010}
J.~Turian, L.~Ratinov, and Y.~Bengio, ``Word representations: a simple and
  general method for semi-supervised learning,'' in \emph{ACM Association for
  Computational Linguistics (ACL 2010)}, Uppsala, Sweden, July 2010, pp.
  384--394.

\bibitem{Krueger_2009}
K.~Krueger and P.~Dayan, ``Flexible shaping: How learning in small steps
  helps,'' \emph{Cognition}, vol. 110, no.~3, pp. 380--394, March 2009.

\bibitem{Sanger_1994}
T.~D. Sanger, ``Neural network learning control of robot manipulators using
  gradually increasing task difficulty,'' \emph{IEEE Transactions on Robotics
  and Automation}, vol.~10, no.~3, pp. 323--333, June 1994.

\bibitem{Baranes_2013}
A.~Baranes and P.~Oudeyer, ``Active learning of inverse models with
  intrinsically motivated goal exploration in robots,'' \emph{Robotics and
  Autonomous Systems}, vol.~61, no.~1, pp. 49--73, January 2013.

\bibitem{Zaremba_2014_2}
W.~Zaremba and I.~Sutskever, ``Learning to execute,'' \emph{ArXiv e-prints
  (arXiv:1410.4615)}, October 2014.

\bibitem{Bengio_2009_2}
Y.~Y.~Bengio, J.~Louradour, R.~Collobert, and J.~Weston, ``Curriculum
  learning,'' in \emph{International Conference on Machine Learning (ICML
  2009)}, Montreal, QC, Canada, June 2009, pp. 41--48.

\bibitem{Busso_2017}
C.~Busso, S.~Parthasarathy, A.~Burmania, M.~AbdelWahab, N.~Sadoughi, and
  E.~{Mower Provost}, ``{MSP-IMPROV}: An acted corpus of dyadic interactions to
  study emotion perception,'' \emph{IEEE Transactions on Affective Computing},
  vol.~8, no.~1, pp. 67--80, January-March 2017.

\bibitem{Zhou_2014}
D.~Zhou, Q.~Liu, J.~Platt, and C.~Meek, ``Aggregating ordinal labels from
  crowds by minimax conditional entropy,'' in \emph{International Conference on
  Machine Learning (ICML 2014)}, Beijing, China, June 2014, pp. 262--270.

\bibitem{Zhou_2015}
D.~Zhou, Q.~Liu, J.~Platt, C.~Meek, and N.~Shah, ``Regularized minimax
  conditional entropy for crowdsourcing,'' \emph{ArXiv e-prints
  (arXiv:1503.07240)}, vol. abs/1503.07240, pp. 1--31, March 2015.

\bibitem{Petrushin_1999}
V.~Petrushin, ``Emotion in speech: Recognition and application to call
  centers,'' in \emph{Proceedings of the Artificial Neural Networks in
  Engineering (ANNIE 1999)}, St. Louis, MO, November 1999.

\bibitem{Dellaert_1996}
F.~Dellaert, T.~Polzin, and A.~Waibel, ``Recognizing emotion in speech,'' in
  \emph{International Conference on Spoken Language (ICSLP 1996)}, vol.~3,
  Philadelphia, PA, USA, October 1996, pp. 1970--1973.

\bibitem{Busso_2004}
C.~Busso, Z.~Deng, S.~Yildirim, M.~Bulut, C.~Lee, A.~Kazemzadeh, S.~Lee,
  U.~Neumann, and S.~Narayanan, ``Analysis of emotion recognition using facial
  expressions, speech and multimodal information,'' in \emph{Sixth
  International Conference on Multimodal Interfaces ICMI 2004}.\hskip 1em plus
  0.5em minus 0.4em\relax State College, PA: ACM Press, October 2004, pp.
  205--211.

\bibitem{Batliner_2000}
A.~Batliner, K.~Fischer, R.~Huber, J.~Spilker, and E.~N\"oth, ``Desperately
  seeking emotions or: actors, wizards and human beings,'' in \emph{ISCA
  Tutorial and Research Workshop (ITRW) on Speech and Emotion}, Newcastle,
  Northern Ireland, UK, September 2000, pp. 195--200.

\bibitem{Devillers_2005}
L.~Devillers, L.~Vidrascu, and L.~Lamel, ``Challenges in real-life emotion
  annotation and machine learning based detection,'' \emph{Neural Networks},
  vol.~18, no.~4, pp. 407--422, May 2005.

\bibitem{Busso_2008_5}
C.~Busso, M.~Bulut, C.~Lee, A.~Kazemzadeh, E.~Mower, S.~Kim, J.~Chang, S.~Lee,
  and S.~Narayanan, ``{IEMOCAP}: Interactive emotional dyadic motion capture
  database,'' \emph{Journal of Language Resources and Evaluation}, vol.~42,
  no.~4, pp. 335--359, December 2008.

\bibitem{Devillers_2006}
L.~Devillers and L.~Vidrascu, ``Real-life emotions detection with lexical and
  paralinguistic cues on human-human call center dialogs,'' in
  \emph{Interspeech - International Conference on Spoken Language (ICSLP)},
  Pittsburgh, PA, USA, September 2006, pp. 801--804.

\bibitem{Grimm_2008}
M.~Grimm, K.~Kroschel, and S.~Narayanan, ``The {V}era {am} {M}ittag {G}erman
  audio-visual emotional speech database,'' in \emph{IEEE International
  Conference on Multimedia and Expo (ICME 2008)}, Hannover, Germany, June 2008,
  pp. 865--868.

\bibitem{Cowie_2000}
R.~Cowie, E.~Douglas-Cowie, S.~Savvidou, E.~McMahon, M.~Sawey, and
  M.~Schr\"oder, ``{'FEELTRACE'}: An instrument for recording perceived emotion
  in real time,'' in \emph{ISCA Tutorial and Research Workshop (ITRW) on Speech
  and Emotion}.\hskip 1em plus 0.5em minus 0.4em\relax Newcastle, Northern
  Ireland, UK: ISCA, September 2000, pp. 19--24.

\bibitem{Burmania_2016_2}
A.~Burmania, S.~Parthasarathy, and C.~Busso, ``Increasing the reliability of
  crowdsourcing evaluations using online quality assessment,'' \emph{IEEE
  Transactions on Affective Computing}, vol.~7, no.~4, pp. 374--388,
  October-December 2016.

\bibitem{Lotfian_2017}
R.~Lotfian and C.~Busso, ``Formulating emotion perception as a probabilistic
  model with application to categorical emotion classification,'' in
  \emph{International Conference on Affective Computing and Intelligent
  Interaction (ACII 2017)}, San Antonio, TX, USA, October 2017, pp. 415--420.

\bibitem{Fayek_2016}
H.~Fayek, M.~Lech, and L.~Cavedon, ``On the correlation and transferability of
  features between automatic speech recognition and speech emotion
  recognition,'' in \emph{Interspeech 2016}, San Francisco, CA, USA, September
  2016, pp. 3618--3622.

\bibitem{Lee_2011}
C.-C. Lee, E.~Mower, C.~Busso, S.~Lee, and S.~Narayanan, ``Emotion recognition
  using a hierarchical binary decision tree approach,'' \emph{Speech
  Communication}, vol.~53, no. 9-10, pp. 1162--1171, November-December 2011.

\bibitem{Lee_2009}
------, ``Emotion recognition using a hierarchical binary decision tree
  approach,'' in \emph{Interspeech 2009}, Brighton, UK, September 2009, pp.
  320--323.

\bibitem{Graves_2017}
A.~Graves, M.~Bellemare, J.~Menick, R.~Munos, and K.~Kavukcuoglu, ``Automated
  curriculum learning for neural networks,'' in \emph{International Conference
  on Machine Learning (ICML 2017)}, Sydney, Australia, August 2017, pp. 1--10.

\bibitem{Gui_2017}
L.~Gui, T.~Baltru{\v{s}}aitis, and L.~Morency, ``Curriculum learning for facial
  expression recognition,'' in \emph{IEEE International Conference on Automatic
  Face and Gesture Recognition (FG 2017)}, Washington, DC, USA, May-June 2017,
  pp. 505--511.

\bibitem{Steidl_2005}
S.~Steidl, M.~Levit, A.~Batliner, E.~N\"oth, and H.~Niemann, ````{Of} all
  things the measure is man'' automatic classification of emotions and
  inter-labeler consistency,'' in \emph{International Conference on Acoustics,
  Speech, and Signal Processing (ICASSP 2005)}, vol.~1, Philadelphia, PA, USA,
  March 2005, pp. 317--320.

\bibitem{Dawid_1979}
A.~Dawid and A.~Skene, ``Maximum likelihood estimation of observer error-rates
  using the {EM} algorithm,'' \emph{Journal of the Royal Statistical Society.
  Series C (Applied Statistics)}, vol.~28, no.~1, pp. 20--28, 1979.

\bibitem{Marsella_2010}
S.~Marsella, J.~Gratch, and P.~Petta, ``Computational models of emotion,'' in
  \emph{A Blueprint for Affective Computing-A sourcebook and manual},
  K.~Scherer, T.~B\"{a}nziger, and E.~Roesch, Eds.\hskip 1em plus 0.5em minus
  0.4em\relax New York, NY: Oxford University Press, November 2010, pp. 21--46.

\bibitem{Zeidner_2003}
M.~Zeidner, G.Matthews, R.~Roberts, and C.~MacCann, ``Development of emotional
  intelligence: Towards a multi-level investment model,'' \emph{Human
  development}, vol.~46, no. 2-3, pp. 69--96, March-June 2003.

\bibitem{Volling_2002}
B.~Volling, N.~McElwain, P.~Notaro, and C.~Herrera, ``Parents' emotional
  availability and infant emotional competence: Predictors of parent-infant
  attachment and emerging self-regulation,'' \emph{Journal of Family
  Psychology}, vol.~16, no.~4, pp. 447--465, 2002.

\bibitem{Mayer_1999}
J.~Mayer, D.~Caruso, and P.~Salovey, ``Emotional intelligence meets traditional
  standards for an intelligence,'' \emph{Intelligence}, vol.~27, no.~4, pp.
  267--298, December 1999.

\bibitem{Mariooryad_2017}
S.~Mariooryad and C.~Busso, ``The cost of dichotomizing continuous labels for
  binary classification problems: Deriving a {Bayesian}-optimal classifier,''
  \emph{IEEE Transactions on Affective Computing}, vol.~8, no.~1, pp. 119--130,
  January-March 2017.

\bibitem{Schuller_2011_3}
B.~Schuller, M.~Valstar, F.~Eyben, G.~McKeown, R.~Cowie, and M.~Pantic,
  ``{AVEC} 2011- the first international audio/visual emotion challenge,'' in
  \emph{Affective Computing and Intelligent Interaction (ACII 2011)}, ser.
  Lecture Notes in Computer Science, S.~D'Mello, A.~Graesser, B.~Schuller, and
  J.-C. Martin, Eds.\hskip 1em plus 0.5em minus 0.4em\relax Memphis, TN, USA:
  Springer Berlin / Heidelberg, October 2011, vol. 6975/2011, pp. 415--424.

\bibitem{Wollmer_2009}
M.~W\"ollmer, F.~Eyben, B.~Schuller, E.~Douglas-Cowie, and R.~Cowie,
  ``Data-driven clustering in emotional space for affect recognition using
  discriminatively trained {LSTM} networks,'' in \emph{Interspeech 2009},
  Brighton, UK, September 2009, pp. 1595--1598.

\bibitem{Rahman_2012}
T.~Rahman and C.~Busso, ``A personalized emotion recognition system using an
  unsupervised feature adaptation scheme,'' in \emph{International Conference
  on Acoustics, Speech, and Signal Processing (ICASSP 2012)}, Kyoto, Japan,
  March 2012, pp. 5117--5120.

\bibitem{Geifman_2017}
Y.~Geifman and R.~El-Yaniv, ``Selective classification for deep neural
  networks,'' in \emph{In Advances in Neural Information Processing Systems
  (NIPS 2017)}, Long Beach, CA, USA, December 2017, pp. 4878--4887.

\bibitem{Lord_1980}
F.~Lord, \emph{Applications of item response theory to practical testing
  problems}.\hskip 1em plus 0.5em minus 0.4em\relax Mahwah, NJ, USA: Lawrence
  Erlbaum Associates, July 1980.

\bibitem{Lotfian_201x}
R.~Lotfian and C.~Busso, ``Building naturalistic emotionally balanced speech
  corpus by retrieving emotional speech from existing podcast recordings,''
  \emph{IEEE Transactions on Affective Computing}, vol. To appear, 2018.

\bibitem{Mariooryad_2014_3}
S.~Mariooryad, R.~Lotfian, and C.~Busso, ``Building a naturalistic emotional
  speech corpus by retrieving expressive behaviors from existing speech
  corpora,'' in \emph{Interspeech 2014}, Singapore, September 2014, pp.
  238--242.

\bibitem{Schuller_2013}
B.~Schuller, S.~Steidl, A.~Batliner, A.~Vinciarelli, K.~Scherer, F.~Ringeval,
  M.~Chetouani, F.~Weninger, F.~Eyben, E.~Marchi, M.~Mortillaro, H.~Salamin,
  A.~Polychroniou, F.~Valente, and S.~Kim, ``The {INTERSPEECH} 2013
  computational paralinguistics challenge: Social signals, conflict, emotion,
  autism,'' in \emph{Interspeech 2013}, Lyon, France, August 2013, pp.
  148--152.

\bibitem{Eyben_2010_2}
F.~Eyben, M.~W\"{o}llmer, and B.~Schuller, ``{OpenSMILE}: the {Munich}
  versatile and fast open-source audio feature extractor,'' in \emph{ACM
  International conference on Multimedia (MM 2010)}, Florence, Italy, October
  2010, pp. 1459--1462.

\bibitem{Kingma_2014_2}
D.~Kingma and J.~Ba, ``Adam: A method for stochastic optimization,'' in
  \emph{International Conference on Learning Representations}, San Diego, CA,
  USA, May 2015, pp. 1--13.

\bibitem{Valstar_2016}
M.~Valstar, J.~Gratch, B.~Schuller, F.~Ringeval, D.~Lalanne, M.~{Torres
  Torres}, S.~Scherer, G.~Stratou, R.~Cowie, and M.~Pantic, ``{AVEC} 2016:
  Depression, mood, and emotion recognition workshop and challenge,'' in
  \emph{International Workshop on Audio/Visual Emotion Challenge}, Amsterdam,
  The Netherlands, October 2016, pp. 3--10.

\bibitem{Maaten_2008}
L.~{van der Maaten} and G.~Hinton, ``Visualizing data using {t-SNE},''
  \emph{Journal of Machine Learning Research}, vol.~9, pp. 2579--2605, November
  2008.

\bibitem{Abdelwahab_2017_2}
M.~Abdelwahab and C.~Busso, ``Incremental adaptation using active learning for
  acoustic emotion recognition,'' in \emph{IEEE International Conference on
  Acoustics, Speech and Signal Processing (ICASSP 2017)}, New Orleans, LA, USA,
  March 2017, pp. 5160--5164.

\end{thebibliography}

\vspace{-1.3cm}
\begin{IEEEbiography}[{\includegraphics[width=1in,height=1.25in,clip,keepaspectratio]{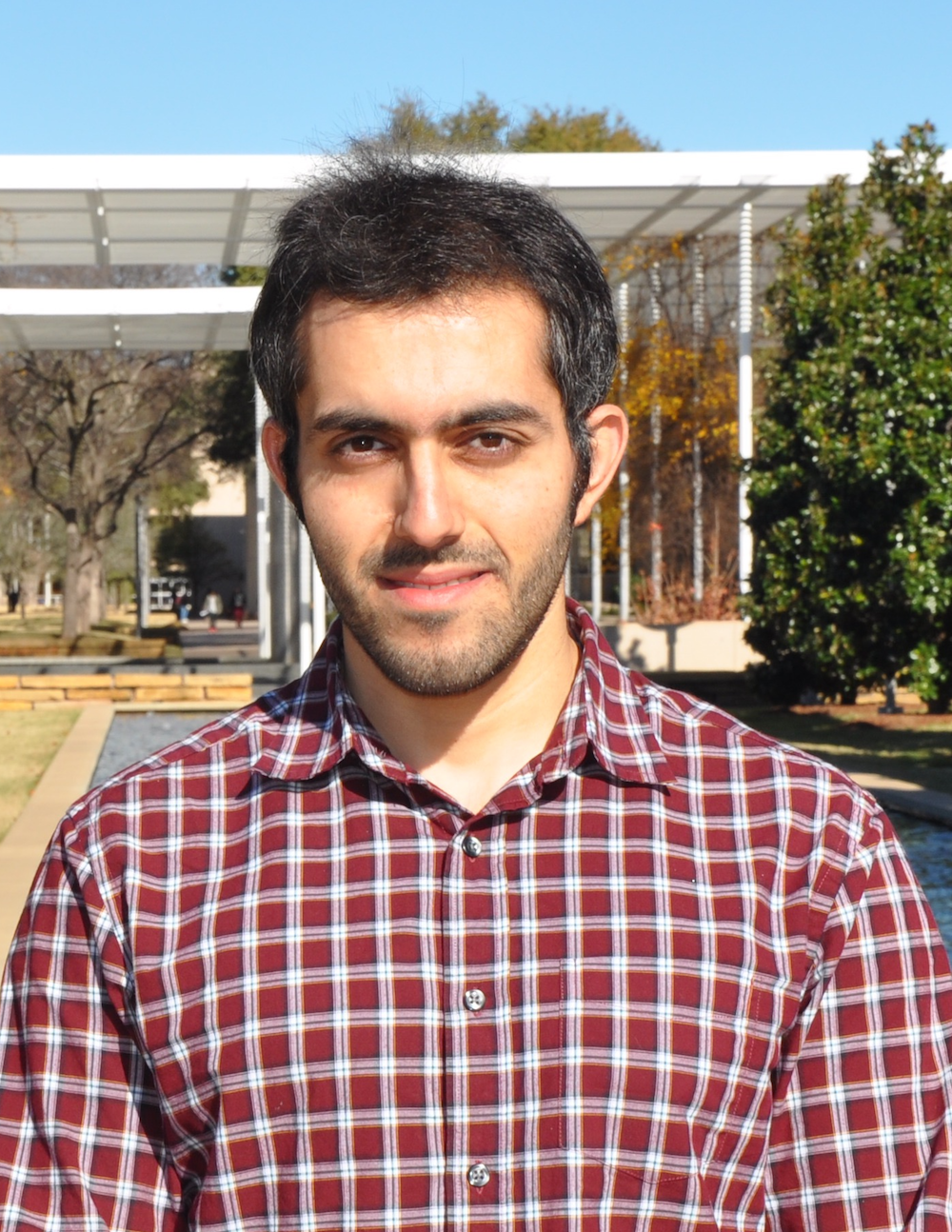}}]{Reza Lotfian} (SM'17) received his BS degree (2006) with high honors in Electrical Engineering from the Department of Electrical Engineering, Amirkabir University, Tehran, Iran and MS degree (2010) in Electrical Engineering from the Sharif University (SUT), Tehran, Iran. He is currently pursuing his Ph.D. degree in the Electrical and Computer Engineering at the University of Texas at Dallas (UTD), Richardson, Texas, USA. He joined the Multimodal Signal Processing (MSP) laboratory in 2013. His research interest includes the area of speech signal processing, affective computing, human machine interaction, and machine learning.
\end{IEEEbiography}

\vspace{-1.0cm}
\begin{IEEEbiography}[{\includegraphics[width=1in,height=1.25in,clip,keepaspectratio]{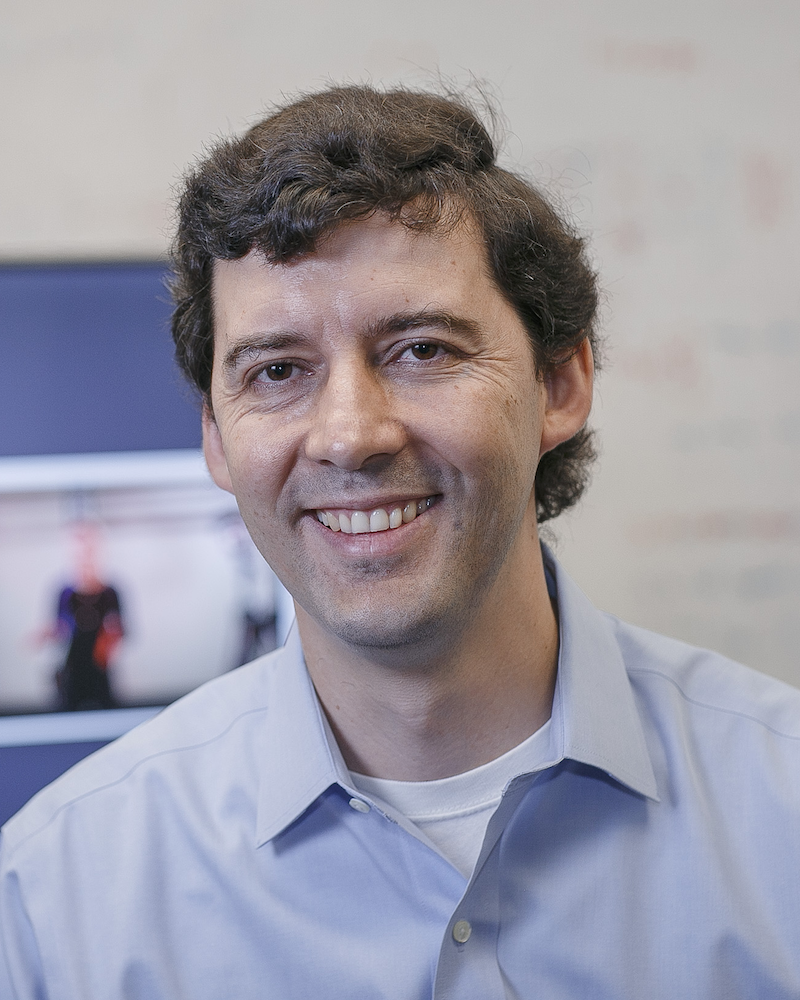}}]{Carlos Busso} 
(S'02-M'09-SM'13) received the BS and MS degrees with high honors in electrical engineering from the University of Chile, Santiago, Chile, in 2000 and 2003, respectively, and the PhD degree (2008) in electrical engineering from the University of Southern California (USC), Los Angeles, in 2008. He is an associate professor at the Electrical Engineering Department of The University of Texas at Dallas (UTD). He was selected by the School of Engineering of Chile as the best electrical engineer graduated in 2003 across Chilean universities. At USC, he received a provost doctoral fellowship from 2003 to 2005 and a fellowship in Digital Scholarship from 2007 to 2008. At UTD, he leads the Multimodal Signal Processing (MSP) laboratory [http://msp.utdallas.edu]. He is a recipient of an NSF CAREER Award. In 2014, he received the ICMI Ten-Year Technical Impact Award. In 2015, his student received the third prize IEEE ITSS Best Dissertation Award (N. Li). He also received the Hewlett Packard Best Paper Award at the IEEE ICME 2011 (with J. Jain), and the Best Paper Award at the AAAC ACII 2017 (with Yannakakis and Cowie). He is the co-author of the winner paper of the Classifier Sub-Challenge event at the Interspeech 2009 emotion challenge. His research interests include digital signal processing, speech and video processing, and multimodal interfaces. His current research includes the broad areas of affective computing, multimodal human-machine interfaces, modeling and synthesis of verbal and nonverbal behaviors, sensing human interaction, in-vehicle active safety system, and machine learning methods for multimodal processing. He was the general chair of ACII 2017. He is a member of ISCA, AAAC, and ACM, and a senior member of the IEEE. \end{IEEEbiography}




\end{document}